\documentclass[lettersize,journal]{IEEEtran}
\usepackage{amsmath,amsfonts}
\usepackage{algorithmic}
\usepackage{algorithm}
\usepackage{array}
\usepackage[caption=false,font=normalsize,labelfont=sf,textfont=sf]{subfig}
\usepackage{textcomp}
\usepackage{stfloats}
\usepackage{url}
\usepackage{verbatim}
\usepackage{graphicx}
\usepackage{cite}
\usepackage{tikz}
\usepackage{soul}
\usepackage{color, xcolor}
\usepackage{amsmath,amssymb,amsthm}
\newtheorem{definition}{Definition}[section]
\newtheorem{theorem}{Theorem}[section]

\usepackage{graphicx}
\usepackage{algorithmic}
\usepackage{algorithm}
\usepackage{textcomp}
\usepackage{soul}
\usepackage{array}
\usepackage{mdframed}
\usepackage{subfig}
\usepackage{bbding}
\usepackage{pifont}
\usepackage{amsmath}
\usepackage{amssymb}

\usepackage{booktabs}
\usepackage{multirow}

\usepackage{subcaption}
\begin{document}

\title{
SLVC-DIDA: Signature-less Verifiable Credential-based Issuer-hiding Authentication for Decentralized Identity}

\author{Tianxiu~Xie,
Keke~Gai,~\IEEEmembership{Senior Member,~IEEE,}
Jing~Yu,~\IEEEmembership{Senior Member,~IEEE,}
Liehuang~Zhu,~\IEEEmembership{Senior Member,~IEEE}
~and~Bin~Xiao,~\IEEEmembership{Fellow,~IEEE}
\thanks{T. Xie, K. Gai and L. Zhu are with the School of Cyberspace Science and Technology, Beijing Institute of Technology, Beijing, China, 100081, \{3120215672, gaikeke, liehuangz\}@bit.edu.cn.}
\thanks{J. Yu is with the School of Information Engineering, Minzu University of China, Beijing, China. (E-mail: jing.yu@muc.edu.cn).}
\thanks{B. Xiao is with Department of Computing, The Hong Kong Polytechnic University, Hong Kong, China. Email: cs-bxiao@comp.polyu.edu.hk.}
\thanks{This work is partially supported by the National Key Research and Development Program of China (Grant No. 2021YFB2701300), National Natural Science Foundation of China (Grant No. 62372044).
}
\thanks{Keke Gai is the corresponding author (gaikeke@bit.edu.cn).}
\thanks{Manuscript received Month Date, 202X.}}

\markboth{Journal of \LaTeX\ Class Files,~Vol.~XX, No.~XX, Month~202X}%
{Shell \MakeLowercase{\textit{et al.}}: A Sample Article Using IEEEtran.cls for IEEE Journals}


\maketitle

\begin{abstract}
As an emerging paradigm in digital identity, Decentralized Identity (DID) appears advantages over traditional identity management methods in a variety of aspects, e.g., enhancing user-centric online services and ensuring complete user autonomy and control. 
Verifiable Credential (VC) techniques are used to facilitate decentralized DID-based access control across multiple entities. 
However, existing DID schemes generally rely on a distributed public key infrastructure that also causes challenges, such as context information deduction, key exposure, and issuer data leakage. 
To address the issues above, this paper proposes a issuer-hiding and privacy-preserving DID multi-party authentication model with a signature-less VC scheme, named  SLVC-DIDA, for the first time. 
Our proposed scheme avoids the dependence on signing keys by employing hashing and issuer membership proofs, which supports universal zero-knowledge multi-party DID authentications, eliminating additional technical integrations. 
We adopt a novel zero-knowledge circuit to maintain the anonymity of the issuer set, thereby enabling public verification while safeguarding the privacy of identity attributes via a Merkle tree-based VC list. 
Furthermore, by eliminating reliance on a Public Key Infrastructure (PKI),  SLVC-DIDA enables decentralized and self-sovereign DID authentication. 
Our experiments further evaluate the effectiveness and practicality of SLVC-DIDA.


\end{abstract}


\section{Introduction}

Digital identity is a fundamental concept of user-centric online services, which is used to uniquely identify an entity's subset of the entire attribute set (e.g., name, age, and post codes) in contemporary digital age. 
Traditional centralized identitiy framework arises security and privacy concerns, since users lack controls over their own identities such that personal data governance and usage are restricted. 
When considering the complexity of modern network applications, the centralized setting also brings the restriction of scalability, single points of failure issue, and interoperability obstacle. 
Decentralized identity is deemed to be a new paradigm for replacing traditional centralized identity governance, promoting a higher-level privacy-preserving and user-centric approach to authentication and authorization \cite{baldimtsi2024zklogin,maram2021candid, peng2024vdid}. 
World Wide Web Consortium (W3C) has standardized and formalized Decentralized Identifier (DID) and Verifiable Credential (VC) for decentralized identity.
Specifically, a DID uniquely identifies an entity's identity and supports proof of ownership of identity attribute data. 
As an alternative to digital certificates, a VC facilitates decentralized identity authentication by establishing an extensive digital identity trust across multiple network services.



However, most existing DID and VC authentication schemes \cite{mazzocca2024evoke,yin2022smartdid,maram2021candid, deng2024futuredid,yin2025dp,yin2025didtrust,liu2025fully} that widely employ digital signatures exhibit a security vulnerability to issuer signing key compromise.
Specifically, several signature-based VC issuance methods \cite{maram2021candid, alangot2022decentralized} still rely on the Public Key Infrastructure (PKI), where Identity Provider (IdP) and Certificate Authority (CA) retain control over signing key management, as well as govern digital identities and encryption operations required for VC issuance. 
The challenging issue is that issuers maybe not competent to keeping signing keys, for instance, due to lack of qualified security infrastructure for governing and defending keys in practice. 
That is to say, adversaries can not only forge legitimate VCs with specific identity attributes but also hide the malicious behaviors, when issuers/signing keys are compromised.


Although current blockchain-based DID authentication solutions \cite{liu2025fully,deng2024futuredid,yin2025dp,yin2025didtrust} address vulnerabilities related to issuer signing keys by decentralizing signature authority through additional committees and distributed key generation protocols, these approaches introduce privacy risks due to potential identity data leakage.
Crucially, the VC issuance process requires users to disclose authentic identity attributes in plaintext to either issuers or committee members (e.g., banks, schools, and medical institutions).
Such requirement fundamentally compromises user sovereignty over personal data, as issuers or committee members gain persistent access to and storage capabilities for sensitive identity information.
Moreover, the VCs in these solutions contains an issuer's signatures that can be validated by the verifier using issuer's public key. 
The verifier can potentially infer a user's identity attributes based on the corresponding issuer in specific scenarios, since each VC is linked to a specific issuer and offers substantial contextual information \cite{connolly2022improved, mir2023aggregate}.
In essence, the exposure of issuer information also poses significant privacy challenges in DID authentication processes.
Notably, some recent independent works consider the hidden characteristics of the issuer and introduce a property called issuer-hiding \cite{begum2023issuer, rosenberg2023zk, bobolz2021issuer, shi2023double}.

 \begin{figure}[!t]
\centerline{\includegraphics[width=1.0\linewidth]{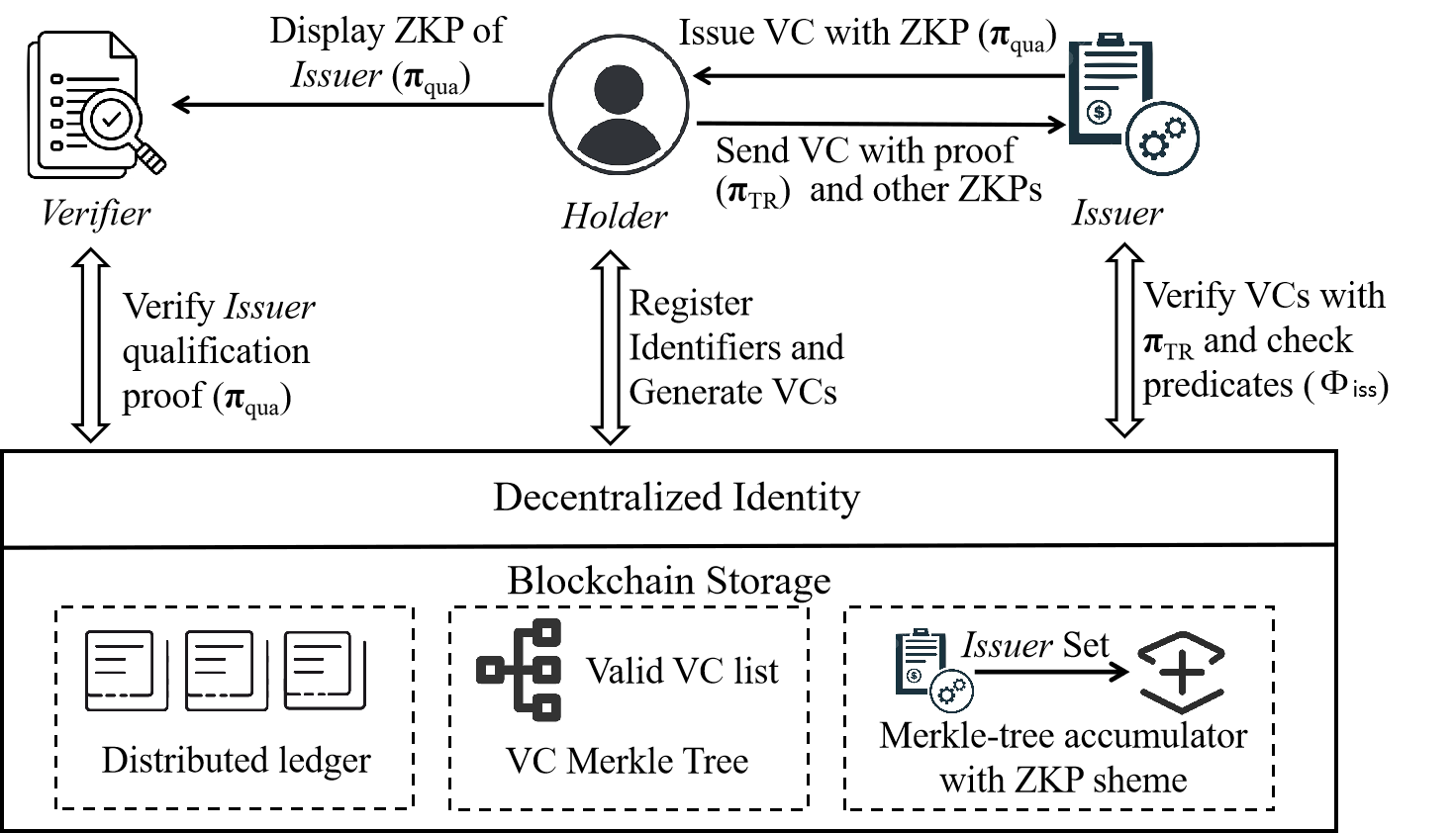}}
 \caption{The high level architecture of our SLVC-DIDA model.} \label{fig:Work}
 \end{figure}

To address the above challenges, this paper proposes the first \textit{\underline{S}ignature-\underline{l}ess \underline{V}erifiable \underline{C}redential-based \underline{DID} \underline{A}uthentication} (SLVC-DIDA) model that achieves issuer-hiding and privacy-preserving decentralized identity verification.
Fig. \ref{fig:Work} illustrates the high-level architecture of SLVC-DIDA.
Our model follows the W3C definitions for \textit{Issuer}, \textit{Holder}, and \textit{Verifier}, with each entity uniquely identified by a DID.
Specifically, inspired by decentralized anonymous credentials \cite{garman2013decentralized,rosenberg2023zk,rathee2022zebra}, VCs are stored and governed on the blockchain via a Merkle-tree-based list structure, ensuring decentralized operation. 
By replacing trust in signing key security with a blockchain-stored VC registry, SLVC-DIDA only need to maintain the complete Merkle trees of VCs to effectively prevent key exposure.

Furthermore, privacy protection for the \textit{Holder} and \textit{Issuer} is realized through our signature-less VC implemented with Zero-Knowledge Proof (ZKP) techniques and commitment schemes.
In SLVC-DIDA, VCs are generated by the \textit{Holder} as commitments to identity attributes and stored on the blockchain. 
During credential application, the \textit{Holder} provides a Merkle-tree membership proof ($\boldsymbol{\pi}_{\mathtt{TR}}$) and attribute commitments (i.e., VCs) instead of revealing plaintext identity attributes to the \textit{Issuer}. 
Despite avoiding identity attribute exposure, valid signature-less VC must undergo \textit{Issuer} verification before issuance. 
Leveraging succinct Non-Interactive Zero-Knowledge (NIZK) arguments of proofs, the \textit{Issuer} checks whether VCs satisfy issuance predicates ($\Phi_{\mathsf{iss}}$) while preserving zero-knowledge properties.
Valid VCs are then issued using the identifiers of the \textit{Holder} and the \textit{Issuer}, with each VC cryptographically bound to its specific \textit{Issuer} through configured randomness rather than public-key signatures.
Critically, we adopt a ZKP scheme that applies another Merkle-tree accumulator with \textit{Issuer} qualification proof ($\boldsymbol{\pi}_{S}$) to certify each valid VC derives from a legitimate \textit{Issuer}. 
While the VC list and the \textit{Issuer} set are stored in two separate on-chain Merkle trees, only the corresponding Merkle roots are publicly exposed.
Finally, for signature-less VC verification, the \textit{Verifier} only requires the qualification ZKP ($\boldsymbol{\pi}_{S}$) of the corresponding legitimate \textit{Issuer} linked to the valid VC, rather than directly obtaining the \textit{Issuer}'s public key.
Consequently, SLVC-DIDA enables privacy-preserving and issuer-hiding identity assertions with signature-less VCs, eliminating dependence on PKI-based signing keys for distributed VC issuance.

The contributions of this paper are outlined as follows:
\begin{enumerate}
    \item Differing from existing DID authentication methods, we propose the first signature-less VC scheme for secure and effective identity authentication in distributed issuance. 
    Our model eliminates reliance on PKI-based digital signatures via blockchain technology, removing the need to trust signing key security. 
    Furthermore, the signature-less VC provides correct identity assertions with full privacy through proofs of legitimate issuer qualification and valid VC membership, rather than redesigning proof systems over issuer-specific public key encryption schemes.
    \item Compared to VC schemes where issuers access user attributes, our SLVC-DIDA enables users to independently generate their own VCs while providing only cryptographic commitments of identity attributes for issuer verification. 
    Through ZKP and blockchain, SLVC-DIDA ensures complete identity confidentiality against issuers/verifiers, delivering end-to-end DID authentication.
    Crucially, all users in SLVC-DIDA maintain complete control over their decentralized identities, including operations of governing and sharing on their VCs.



    \item Our proposed SLVC-DIDA model ensures issuer-hiding while maintaining DID authentication integrity through ZKP and Merkle-tree accumulator.
    Rather than directly accessing issuer public keys, SLVC-DIDA leverages a Merkle-tree-based membership proof to verify whether the VC originate from a legitimate issuer.
    Moreover, the issuer generates a one-time randomness to cryptographically bind VCs, providing one-way verifiability through the hash value of the randomness. 
    This process effectively prevents disclosure of issuer contextual information during VC issuance.

\end{enumerate}

The organization of this paper follows the order below.
Related work is given in Section \ref{sec:relatedwork}.
The background and preliminaries of DID is given in Section \ref{sec:pre}.
Sections \ref{sec:mod} and \ref{sec:method} present the design of the proposed model and explanations of the constructions of the proposed method, respectively. 
Sections \ref{sec:exp} provide analysis of the proposed model and experiment evaluations with findings. 
Finally, we draw our conclusions in Section \ref{sec:coc}.
\section{Related Work}
\label{sec:relatedwork}





\textbf{DID Framework.}
The development of DID framework has been addressed by a few recent studies that explored the technical capabilities, such as scalability and security \cite{javed2021health,xie2023cross,xiong2023bdim}. 
For example, Deepak {\em et al.} \cite{maram2021candid} developed a DID framework, called CanDID, that used web authentication services to address the bootstrapping challenge through a decentralized node committee for seamless credential issuance, so that the capability of users' control over identity data was improved. 
Liu {\em et al.} \cite{liu2024ss} conducted a method of establishing trust via appling leader sharding and main chain, which used conventional sharding to manage DID related transactions. 
Some other efforts were made to reduce computation and storage costs, such as using a cryptographic accumulator-based solution to constructing a revocation mechanism for VCs in IoT networks \cite{mazzocca2024evoke}. 
Despite the decentralized nature of DID frameworks, these solutions fail to address privacy protection for both users and issuers.


\textbf{Decentralized Identity Authentication.}
Existing identity authentication schemes are primarily based on cryptographic techniques.
For example, aggregate signature-credential have been probed to secure identity verification by prior studies \cite{mir2023aggregate,hebant2023traceable}. 
Hébant {\em et al.}
\cite{hebant2023traceable} introduced the use of aggregate signatures with randomized tags, allowing specific tracking authorities to perform tracking. 
Mir {\em et al.} 
\cite{mir2023aggregate} employed two signature primitives, aggregated signatures with randomized tags and public keys and aggregate Mercurial signatures, to propose an issuer-hiding multi-authorization anonymous credential scheme.
%
Moreover, blind signatures \cite{hanzlik2023non, karantaidou2024blind} and ring signatures \cite{kasimatis2024did,liang2023lrs_pki} are also used in identity authentication to preserving privacy for users or issuers.
However, the above PKI-based approaches for credential issuance and verification remain vulnerable to risks posed by a single incompetent issuer, including signing key compromise or lower-level security protocol for VC issuance.
Additionally, neither blind nor ring signatures can achieve issuer-hiding and privacy-preserving DID authentication in a single scheme.
In contrast, our signature-less VC achieves both properties through ZKPs.

To decentralize the trust relying on a single issuer, Another type of methods has explored cryptographic credential-based schemes for achieving distributed issuance of DIDs and VCs across multiple authorities, such as threshold signatures \cite{li2024dtacb,mir2023threshold,shi2023threshold,sonnino2018coconut} and distributed key generation \cite{liu2025fully,deng2024futuredid,yin2025dp,yin2025didtrust}. 
Since threshold credentials typically enable users to obtain credentials in a decentralized manner while maintaining their privacy, Li {\em et al.} \cite{li2024dtacb} proposed a dynamic method to support batch display of credentials and proof of the number of credentials, while keeping the user's credential set private.
Sonnino {\em et al.}\cite{sonnino2018coconut} proposed a selective disclosure credential scheme that supports distributed threshold issuance, public and private attributes, re-randomization, and multiple un-linkable selective attribute revelations.
For distributed key generation, DP-DID \cite{yin2025dp} proposes a dynamic committees with batch proactive secret sharing to resistant the mobile adversary attacks.
While the above schemes achieve privacy preservation of user identity attributes toward verifiers, the authentic user data are still disclosed to issuers for VC issuance. 
Furthermore, the issuer-hiding problem remains unaddressed.


\textbf{Privacy-preserving Identity.}
For DID authentication, user identity information must not be disclosed, thus requiring privacy protection for identity data. 
Prior studies \cite{hsieh2024physiological,ye2024securereid, yin2022smartdid, luong2023privacy, yeoh2023fast,jia2025multi} have probed numerous methods to preserve privacy from various perspective. 
In the context of DID, existing methods are involved within two major categories, including attribute-based encryption and ZKP techniques.
For example, Campanelli {\em et al.} \cite{campanelli2022succinct} tried using zkSNARKs and RSA accumulators to prove batch membership and batch updates in zero-knowledge. 
This method supports effective privacy-preserving verification of identity attributes in DID and VC. 
Alangot {\em et al.} \cite{alangot2022decentralized} was an attempt to integrate Verifiable Random Functions (VRF) with blockchain technology, such that it balances auditability and privacy protections. 
Even though the progress is made as above, a critical issue still restricts these methods in DID framework, as the decentralized issuance and distributed management inherent had been rarely addressed. 
\section{Background and Preliminaries}\label{sec:pre}

\textbf{Collision-resistant Hash Function.}
Collision-resistance and one-wayness are the key properties of hash functions.
A typical hash function $\mathsf{CRH}$ involves two Probabilistic Polynomial-time (PPT) algorithms $(\mathsf{CRH.Setup,CRH.Hash})$ as follows. 
\begin{itemize}
    \item $\mathtt{pp_{h}}$ $\gets$ $\mathsf{CRH.Setup}(1^{\lambda})$: Inputs the security parameter $\lambda$ and outputs the public parameters $\mathtt{pp_{h}}$ for collision-resistant hash function.
    \item $\mathtt{h}(m)$ $\gets$ $\mathsf{CRH.Hash}_{\mathtt{pp_{h}}}(m)$: Based on public parameters $\mathtt{pp_{h}}$, $\mathsf{CRH.Hash}(\cdot)$ inputs message $m$ and outputs a short hash value $\mathtt{h}(m)$.
\end{itemize}


\textbf{Cryptographic Accumulator.}
Cryptographic accumulators are used to generate a concise commitment that binds a set of elements together and provides succinct membership proofs for any element within the set. 
These proofs can be publicly verified using the commitment. 
Merkle hash tree is a widely utilized cryptographic accumulator, structured as a binary tree. 
Merkle hash tree consists of six PPT algorithms  $(\mathsf{MK.Setup,}$ $\mathsf{MK.TreeGen,}$ $\mathsf{MK.ComMem,}$ $\mathsf{MK.Verify}, $ $\mathsf{MK.Add},$ $ \mathsf{MK.Delete})$:

\begin{itemize}
    \item $\mathtt{pp_{mk}}$ $\gets$ $\mathsf{MK.Setup}(1^{\lambda})$: Inputs the security parameter $\lambda$ and outputs the public parameters $\mathtt{pp_{mk}}$.


    \item $\mathtt{TR}$ $\gets$ $\mathsf{MK.TreeGen}_{\mathtt{pp_{mk}}}(X)$: Inputs a set $X$ and then outputs the Merkle hash tree $\mathtt{TR}$ with root value $\mathtt{Tr}$, constructed from the elements of $X$.

    \item $\boldsymbol{\pi}$ $\gets$ $\mathsf{MK.ComMem}_{\mathtt{pp_{mk}}}(x,\mathtt{TR})$: 
    On input an element \( x \) and Merkle tree $\mathtt{TR}$, $\mathsf{MK.ComMem}(\cdot)$ outputs a membership proof $\boldsymbol{\pi}$ by computing the authentication path from the root node to the leaf node corresponding to $x$, which proves \( x  \in \mathtt{TR} \).

    \item $\left\{ 0,1 \right\}$ $\gets$ $\mathsf{MK.Verify}_{\mathtt{pp_{mk}}}$ $(x, \boldsymbol{\pi}, \mathtt{Tr})$: 
    On input the membership proof $\boldsymbol{\pi}$, an element \( x \) and Merkle root $\mathtt{Tr}$, it outputs $1$ for \( x  \in \mathtt{TR} \). Otherwise, it outputs $0$.

    \item $\mathtt{TR}'$ $\gets$ $\mathsf{MK.Add}_{\mathtt{pp_{mk}}}(x,\mathtt{TR})$: On input a new element $x$ and original tree $\mathtt{TR}$, $\mathsf{MK.Add}(\cdot)$ adds $x$ to the next free leaf node of $\mathtt{TR}$ and outputs $\mathtt{TR}'$ as the updated Merkle hash tree.

    \item $\mathtt{TR}'$ $\gets$ $\mathsf{MK.Delete}_{\mathtt{pp_{mk}}}(x,\mathtt{TR})$: If $x$ is present in $\mathtt{TR}$, $\mathsf{MK.Delete}(\cdot)$ removes $x$ from $\mathtt{TR}$ and outputs the updated tree $\mathtt{TR}'$.

\end{itemize}




\textbf{Non-Interactive Zero-Knowledge (NIZK) Arguments of Knowledge.}
NIZK is a two-party cryptographic protocol that allows one party (the \textbf{prover}) to prove the correctness of an NP statement to another party (the \textit{Verifier}) without providing any additional information and without any interaction between the two parties. 
The cryptographic primitive of NIZK is a tuple of three PPT algorithms $(\mathsf{NIZK.Setup,}$ $\mathsf{NIZK.Prove,}$ $\mathsf{NIZK.Verify})$ with the following syntax:
\begin{itemize}
    \item $\mathtt{pp_{zk}}$ $\gets$ $\mathsf{NIZK.Setup}(\mathcal{R}, 1^{\lambda})$: On input the security parameter $\lambda$ and the specification of an NP relationship $\mathcal{R}$, $\mathsf{NIZK.Setup}(\cdot)$ outputs the public parameters $\mathtt{pp_{zk}}$ as a relation-specific common reference string (crs).


    \item  $\boldsymbol{\pi}_{zk}$ $\gets$ $\mathsf{NIZK.Prove}_{\mathtt{pp_{zk}}}(x,w)$: Based on $\mathtt{pp_{zk}}$, the algorithm $\mathsf{NIZK.Prove}(\cdot)$ inputs a statement $x$ and a witness $w$ such that the statement-witness pair $(x,w) \in \mathcal{R}$. 
    It outputs the proof $\boldsymbol{\pi}_{zk}$ of the relationship.

    \item $\left\{ 0,1 \right\}$ $\gets$ $\mathsf{NIZK.Verify}_{\mathtt{pp_{zk}}}(\boldsymbol{\pi}_{zk},x)$: On input the statement $x$ and the proof $\boldsymbol{\pi}_{zk}$, $\mathsf{NIZK.Verify}(\cdot)$ outputs a bit value of $0$ or $1$.
    
\end{itemize}




\textbf{Zero-knowledge-Supporting Documentation.}
Rosenberg et al. \cite{rosenberg2023zk} introduce the concept of Zero-Knowledge-Supporting Documentation (ZKSD), a ZKP derived from authoritative identity documents (e.g., e-passports, national IDs).
Without disclosing any sensitive information, ZKSD cryptographically asserts that: (i) attributes in a (non-authoritative) credential match the corresponding data in the original authoritative document, and (ii) the digital signature from the issuing authority on the original document remains valid.
Typically, a ZKSD (denoted as $\mathtt{ZD}$) is generated by two elements: private witness ($w_{\mathtt{ZD}}$), and public auxiliary metadata ($\mathtt{aux}_{\mathtt{ZD}}$).

For example, an age-derived credential is generated from an e-passport, where $w_{\mathtt{ZD}}$ represents the passport's date of birth and signing key while $\mathtt{aux}_{\mathtt{ZD}}$ contains metadata such as current date, authority identifier, and issuance conditions.
Thus, the $\mathtt{ZD}$ generated from $w_{\mathtt{ZD}}$ and $\mathtt{aux}_{\mathtt{ZD}}$ can assert that the age attribute in the derived credential matches that in the passport, and that the passport is issued by the government.



\section{SLVC-DIDA Design}
\label{sec:mod}


\subsection{Threat Model}


\textit{Adversary's Objective:} 
Adversaries may deviate from the VC validation and DID authentication protocols through speculative manipulation for illicit profits, such as inferring the \textit{Holder}'s identity data or illegally obtaining identity authorization.

\textit{Adversary's Knowledge:}
Adversaries have full access to all public inputs of the ZKP circuit as well as all publicly available data on the blockchain, including the randomness ($r_I$) from \textit{Issuer}, the Merkle root ($\mathtt{Tr_{VC}}$) of the VC list, the Merkle root ($\mathtt{Tr_{IS}}$) of legitimate \textit{Issuer} set, and all public parameters ($\mathtt{pp}$), etc.. 

\textit{Adversary's Capability:}
Our SLVC-DIDA model considers static corruption by Probabilistic Polynomial Time (PPT) adversaries who corrupt entities before protocol execution but cannot perform dynamic corruption during runtime. 
Specifically, the malicious behaviors of each entity involved in DID authentication are as follows:

\begin{itemize}
    \item \textit{Misbehaved Holder:} Since the VC is generated by \textit{Holder} in SLVC-DIDA, we assume that \textit{Holders} are arbitrarily malicious and may deviate from the predefined protocols. 
    By seizing control of their identity data, \textit{Holders} prioritize their own interests and generate falsified/misleading VCs to deceive \textit{Issuers} and \textit{Verifiers}, so that they potentially allow unauthorized entities to gain permissions within the network services. 
    \item \textit{Semi-trusted Issuer:} \textit{Issuers} are honest-but-curious. They adhere to the signature-less VC issuance protocols but may attempt to infer the \textit{Holders}' authentic identity data.

    \item \textit{Semi-trusted Verifier:} Similar to \textit{Issuers}, the honest-but-curious \textit{Verifiers} correctly executes the VC verification protocol, but they may attempt to infer the \textit{Holders}' identity data through contextual information, such as the VC content and the \textit{Issuer}'s privacy.

\end{itemize}



\subsection{Design Goals}
SLVC-DIDA with signature-less VC should achieve the following goals for privacy-preserving and issuer-hiding DID authentication:

\begin{itemize}
    \item \textbf{Decentralization.}
    SLVC-DIDA employs blockchain technology to decentralize the entire DID authentication process. 
    This approach distributes VC generation, storage, and verification across the blockchain network rather than relying on CAs and IdPs.

    \item \textbf{Self-sovereign Identity.} 
    SLVC-DIDA should ensure that \textit{Holders} retain full control over their personal data, eliminating the need to disclose any authentic identity attributes to \textit{Issuers} or \textit{Verifiers} during VC issuance and verification.
    Furthermore, the \textit{Holder} can independently generate their own VCs based on a ZKSD.


    \item \textbf{Privacy.} 
    Signature-less VCs must achieve dual objectives: (i) providing correct identity assertions with cryptographic completeness and soundness, and (ii) ensuring privacy-preserving and issuer-hiding properties.
    Specifically, SLVC-DIDA should ensure that all VC with qualification ZKPs ($\boldsymbol{\pi}_{S}$) and membership proofs ($\boldsymbol{\pi}_{\mathtt{VC}}$) reveal zero knowledge about the \textit{Holder}'s specific identity attributes to adversaries.
    Additionally, the qualification ZKP ($\boldsymbol{\pi}_{S}$) should support zero-knowledge verification that proves the valid VC is issued from a legitimate \textit{Issuer} without revealing the \textit{Issuer}'s identifier ($\mathtt{DID^{I}}$).


    

    

    
\end{itemize}


\subsection{Model Design}

The SLVC-DIDA model comprises four core entities: \textit{Holder}, \textit{Issuer}, \textit{Verifier}, and blockchain. 
The first three entities are uniquely identified by decentralized identifiers $\mathtt{DID^{H}}$, $\mathtt{DID^{I}}$, and $\mathtt{DID^{V}}$, respectively. 
Before initiating the DID authentication process, all VCs generated by the \textit{Holder} are stored on the blockchain as a Merkle tree ($\mathtt{TR_{VC}}$ with $\mathtt{pp_{mk}}$), whose root is denoted as $\mathtt{Tr_{VC}}$. 
Meanwhile, the legitimate \textit{Issuer} set ($\mathtt{IS}$) is maintained on-chain as a separate Merkle tree ($\mathtt{TR_{IS}}$ with $\mathtt{pp_{mk}}$), with Merkle root ($\mathtt{Tr_{IS}}$). 
Beyond plaintext attributes, the \textit{Holder} possesses a ZKSD witness ($w_{\mathtt{ZD}}$) and auxiliary metadata ($\mathtt{aux}_{\mathtt{ZD}}$) for valid VC generation.
Additionally, SLVC-DIDA relies on blockchain to ensure trustworthy VC verification and reliable DID authentication through secure consensus mechanisms. 
We assume that the blockchain enables entities to agree on the valid VC list and effectively manage the Merkle-tree accumulator.


\begin{figure}[!t]
\centering
\centerline{\includegraphics[width=1.0\columnwidth]{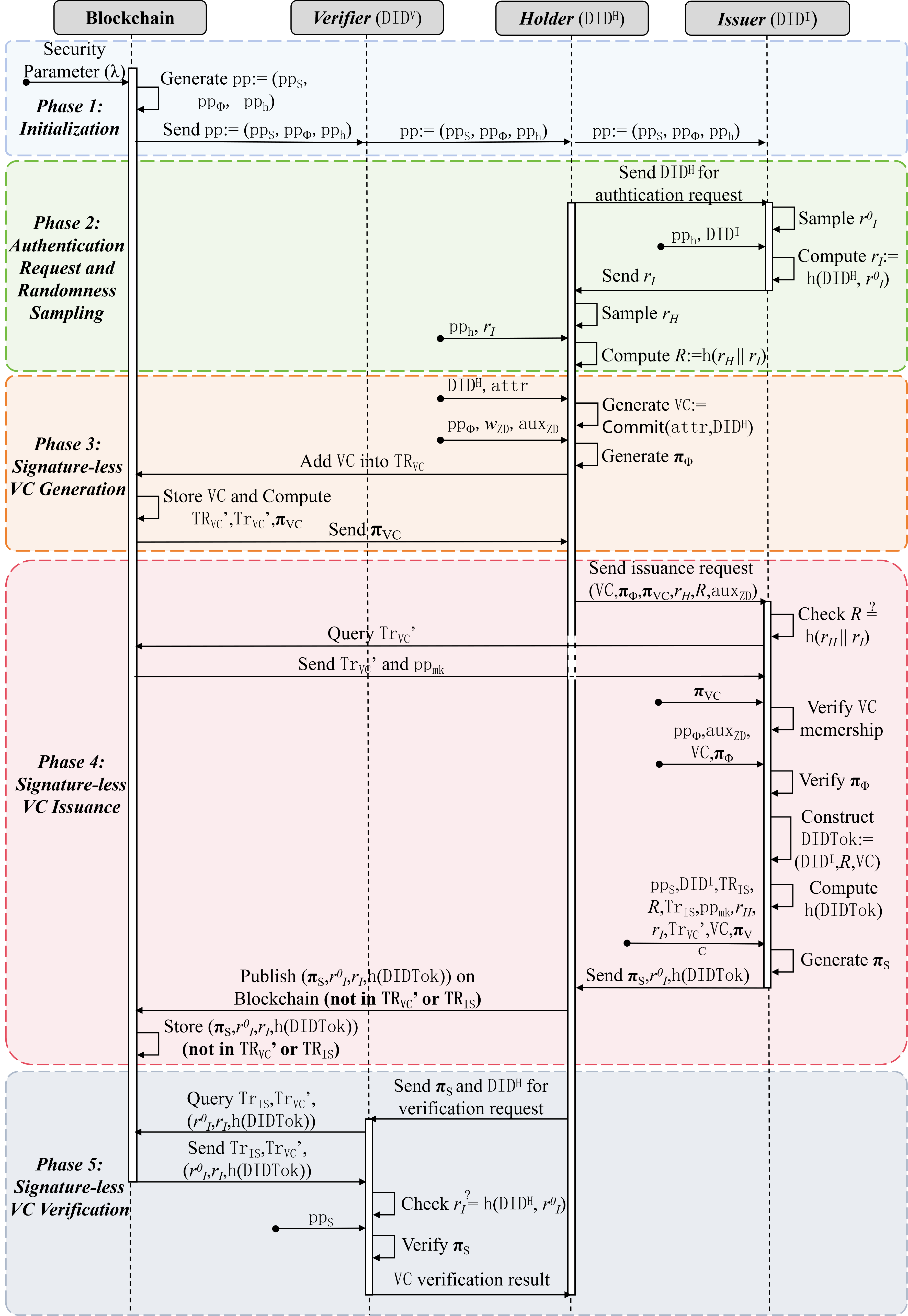}}
\caption{The workflow of DID authentication in SLVC-DIDA.
} \label{fig:Flow}
\end{figure}

SLVC-DIDA employs the signature-less VC to realize DID authentication for the \textit{Holder}. 
Fig. \ref{fig:Flow} shows the work flow of DID authentication in SLVC-DIDA.
The complete DID authentication process consists of five phases: \textit{Initialization}, \textit{Authentication Request and Randomness Sampling}, \textit{Signature-less VC Generation}, \textit{Signature-less VC Issuance}, and \textit{Signature-less VC Verification}. 


\begin{itemize}
    \item \textit{Phase 1: Initialization.}
    SLVC-DIDA generates ZKP public parameters ($\mathtt{pp}$) based on the security parameter ($\lambda$), comprising three parameters: (i) $\mathtt{pp}_{\Phi}$ for checking VC issuance predicate; (ii) $\mathtt{pp}_{S}$ for checking the \textit{Issuer} qualification; and (iii) $\mathtt{pp_{h}}$ for collision-resisitant hash function.
    These parameters ($\mathtt{pp}$) are accessible to all cryptographic algorithms within the SLVC-DIDA model.


    \item \textit{Phase 2: Authentication Request and Randomness Sampling.}
    To prevent replay attacks, SLVC-DIDA establishes session binding for signature-less VCs through a two-party randomness generation process. 
    When the \textit{Issuer} (identified by $\mathtt{DID^{I}}$) receives an authentication request from the \textit{Holder} ($\mathtt{DID^{H}}$), it first samples fresh randomness ($r_I^0$) and computes $r_I := \mathtt{h}(\mathtt{DID^{H}}, r_I^0)$, which is then sent to the \textit{Holder}. 
    The \textit{Holder} then samples its own randomness ($r_H$) and computes $R := \mathtt{h}(r_H \parallel r_I)$, which is used for all subsequent operations within the session, where $\parallel$ denotes concatenation. Then $R$ and $r_H$ is send to the \textit{Issuer} with VC in \textit{Phase 4}.
    Each VC issuance is bound to a specific DID authentication session through the randomness ($r_H$, $r_I^0$). 
    The hash computation for $r_I$ (generated by the \textit{Issuer}) and $R$ (generated by the \textit{Holder}) is shown as follows:
    \begin{align*}
        &r_I := \mathtt{h}(\mathtt{DID^{H}}, r_I^0) \gets \mathsf{CRH.Hash}_{\mathtt{pp_{h}}}(\mathtt{DID^{H}}, r_I^0) \\
        &R := \mathtt{h}(r_H \parallel r_I) \gets \mathsf{CRH.Hash}_{\mathtt{pp_{h}}}(r_H \parallel r_I)
    \end{align*}


    \item \textit{Phase 3: Signature-less VC Generation.}
    The \textit{Holder} commits to its attributes ($\mathtt{attr}$) and identifier ($\mathtt{DID^{H}}$) to generate a signature-less VC, i.e., $\mathtt{VC}:=\mathsf{Commit}(\mathtt{attr}, \mathtt{DID^{H}})$.
    Notably, to prevent falsified/misleading VCs, the \textit{Holder} is required to generate an additional issuance ZKP ($\boldsymbol{\pi}_{\Phi}$) based on ZKSD ($\mathtt{ZD}$) and public parameter ($\mathtt{pp}_{\Phi}$), which correctly asserts that: (i) the $\mathtt{VC}$ is derived from authoritative identity documents, and (ii) the attributes committed in the $\mathtt{VC}$ satisfy the \textit{Issuer}'s issuance predicate ($\Phi_{\mathsf{iss}}$).
    Finally, the generated VC is stored into the on-chain Merkle tree ($\mathtt{TR_{VC}}$) of VC list.
    Blockchain updates the Merkle tree to $\mathtt{TR'}$ with a new root ($\mathtt{Tr_{VC}'}$), while computing the membership proof ($\boldsymbol{\pi}_{\mathtt{VC}}$) for the newly added $\mathtt{VC}$.


    \item \textit{Phase 4: Signature-less VC Issuance.}
    The \textit{Holder} submits an issuance request to the \textit{Issuer}, including the signature-less VC ($\mathtt{VC}$), proofs ($\boldsymbol{\pi}_{\Phi}$, $\boldsymbol{\pi}_{\mathtt{VC}}$), randomness ($r_H$, $R$) and the auxiliary metadata ($\mathtt{aux}_{\mathtt{ZD}}$).
    The \textit{Issuer} first checks if $R$ matches the hash output of the randomness ($r_H$ $\parallel$ $r_I$), i.e., $R$ $\overset{?}{=}$ $\mathtt{h}(r_H \parallel r_I)$.
    Then it verifies the the on-chain membership of $\mathtt{VC}$ via its proof ($\boldsymbol{\pi}_{\mathtt{VC}}$) and the public Merkle root ($\mathtt{Tr_{VC}'}$). 
    Additionally, the \textit{Issuer} verifies the issuance proof ($\boldsymbol{\pi}_{\Phi}$) to certify the correctness of the $\mathtt{VC}$'s identity assertions, i.e., (i) and (ii) in \textit{Phase 2}. 
    Successful verification of both proofs ($\boldsymbol{\pi}_{\mathtt{VC}}$, $\boldsymbol{\pi}_{\Phi}$) establishes the $\mathtt{VC}$'s validity, prompting the \textit{Issuer} to generate a legitimacy attestation.
    Specifically, the \textit{Issuer} with $\mathtt{DID^{I}}$ generates a qualification ZKP ($\boldsymbol{\pi}_{S}$) to assert that: (i) the \textit{Issuer} ($\mathtt{DID^{I}}$) belongs to Merkle-tree accumulator ($\mathtt{TR_{IS}}$) of the legitimate \textit{Issuer} set ($\mathtt{IS}$) 
    and (ii) the valid $\mathtt{VC}$ is properly issued by the corresponding \textit{Issuer} ($\mathtt{DID^{I}}$). 
    In particular, for the ZKP circuit of qualification proof ($\boldsymbol{\pi}_{S}$), the \textit{Issuer} constructs a private DID token ($\mathtt{DIDTok}$) containing hashed randomness ($R$) to cryptographically bind the $\mathtt{VC}$ with $\mathtt{DID^{I}}$.
    The DID token is constructed as follows:
    \begin{align*}
        &\mathtt{DIDTok} := (\mathtt{DID^{I}},\mathtt{VC},R) := (\mathtt{DID^{I}},\mathtt{VC},\mathtt{h}(r_H \parallel r_I))\\
        &\mathtt{h}(\mathtt{DIDTok}) \gets \mathsf{CRH.Hash}_{\mathtt{pp_{h}}}(\mathtt{DIDTok})
    \end{align*}
    The \textit{Issuer} transmits the qualification ZKP ($\boldsymbol{\pi}_{S}$) of $\mathtt{VC}$ back to the \textit{Holder}.
    The \textit{Holder} then publishes the tuple ($\boldsymbol{\pi}_{S}$, $r_I^0$, $r_I$, $\mathtt{h}(\mathtt{DIDTok})$) on the blockchain. 
    We remark that the above tuple tuple is published as public on-chain data, distinct from and independent of the Merkle tree ($\mathtt{TR_{VC}}'$) of the VC list and the Merkle-tree accumulator ($\mathtt{TR_{IS}}$) of the \textit{Issuer} set.


    \item \textit{Phase 5: Signature-less VC Verification.}
    During VC verification, the \textit{Holder} submits its identifier ($\mathtt{DID^{H}}$) and the qualification ZKP ($\boldsymbol{\pi}_{S}$) of $\mathtt{VC}$ to the \textit{Verifier} without disclosing the $\mathtt{VC}$.
    Based on $\boldsymbol{\pi}_{S}$, the \textit{Verifier} retrieves the corresponding tuple ($r_I^0$, $r_I$, $\mathtt{h}(\mathtt{DIDTok})$) from the blockchain and first checks whether $r_I$ $\overset{?}{=}$ $\mathtt{h}(\mathtt{DID^{H}},r_I^0)$ holds.
    Subsequently, the \textit{Verifier} verifies the ZKP ($\boldsymbol{\pi}_{S}$) against both Merkle roots ($\mathtt{Tr_{IS}}$ and $
    \mathtt{Tr_{VC}'}$).
    Finally, the verification result of $\mathtt{VC}$ is return to the \textit{Holder}.

    
\end{itemize}


\section{Signature-less Verifiable Credential}
\label{sec:method}

\subsection{Signature-less VC Construction}
In SLVC-DIDA, the signature-less VC provides three key functionalities for the \textit{Holder}: generation, issuance, and verification. 
The \textit{Holder} can generate a commitment ($\mathtt{VC}$) for its identity attributes and identifier, which is stored in a Merkle tree ($\mathtt{TR_{VC}}$). 
During the issuance process, the \textit{Issuer} validates the \textit{Holder}'s generated VC, then sends a signature-less VC tuple to the \textit{Holder}.
Following the blockchain publication of this tuple, the \textit{Holder} can leverage the signature-less VC to cryptographically prove satisfaction of the issuance predicate/access criteria.
Specifically, a valid signature-less VC involves three proofs: (i) a membership proof ($\boldsymbol{\pi}_{\mathtt{VC}}$) asserting the existence of $\mathtt{VC}$ within the Merkle tree ($\mathtt{TR_{VC}}$), (ii) an issuance ZKP ($\boldsymbol{\pi}_{\Phi}$) certifying that the committed attributes ($\mathtt{attr}$) satisfy the issuance predicate ($\Phi_{\mathsf{iss}}$), and (iii) a qualification ZKP ($\boldsymbol{\pi}_{S}$) verifying the legitimacy of the \textit{Issuer}. 
We note that the membership proof ($\boldsymbol{\pi}_{\mathtt{VC}}$) in SLVC-DIDA does not possess zero-knowledge properties.
The algorithms of signature-less VC is shown as follows:

\begin{itemize}
    \item $\mathtt{pp} \gets \mathsf{Setup}(1^{\lambda})$: This algorithm is execute by the blockchain to generates the public parameters ($\mathtt{pp}$) in the \textit{Phase 1} for all other algorithms.
    
    \item $(\mathtt{VC},\boldsymbol{\pi}_{\mathtt{VC}})$ $\gets$ $\mathsf{GenVC}_{\mathtt{pp}}(\mathtt{attr},\mathtt{DID^{H}},\mathtt{TR_{VC}};\mathtt{pp_{mk}})$: 
    During \textit{Phase 3}, this algorithm is executed by the \textit{Holder} to generate the signature-less VC with its attributes and identifier. Then the new VC is added into the Merkle tree of VC list and returns the corresponding membership proof ($\boldsymbol{\pi}_{\mathtt{VC}}$).
    
    \item $\boldsymbol{\pi}_{\Phi}$ $\gets$ $\mathsf{AuthReq}_{\mathtt{pp}}(\mathtt{VC,\mathtt{aux_{ZD}}},w_{\mathtt{ZD}})$:
    During \textit{Phase 3}, this algorithm is executed by the \textit{Holder} to generate the issuance proof ($\boldsymbol{\pi}_{\Phi}$) of $\mathtt{VC}$ based on the ZKSD ($\mathtt{ZD}$ with $\mathtt{aux_{ZD}}$ and $w_{\mathtt{ZD}}$) under the issuance predicate ($\Phi_{\mathsf{iss}}$). 

    \item $\{ 0,1\}$ $\gets$ $\mathsf{VerifyIss}_{\mathtt{pp}}$ $(\mathtt{pp_{mk}},\mathtt{VC},\boldsymbol{\pi}_{\mathtt{VC}},\boldsymbol{\pi}_{\Phi},\mathtt{Tr_{VC}}',\mathtt{aux_{ZD}},R,$ $r_H,r_I)$: 
    During \textit{Phase 4}, this algorithm is executed by the \textit{Issuer} to determine whether to issue $\mathtt{VC}$. 
    After checking the session binding via the hash ($R$), the \textit{Issuer} verifies the $\mathtt{VC}$ with membership proof ($\boldsymbol{\pi}_{\mathtt{VC}}$) and issuance proof ($\boldsymbol{\pi}_{\Phi}$). If both proofs are correct, the algorithm outputs $1$, approving the $\mathtt{VC}$ for issuance. 
    
    \item 
    $(\boldsymbol{\pi}_{S}, \mathtt{h}(\mathtt{DIDTok}))$ $\gets$ $\mathsf{IssueVC}_{\mathtt{pp}}(\mathtt{VC},$ $\mathtt{DID^{I}},$ $\mathtt{IS},$ $\mathtt{Tr_{IS}},$ $\boldsymbol{\pi}_{\mathtt{VC}},$ $\mathtt{Tr_{VC}}',R,$ $r_H,r_I,\mathtt{pp_{mk}})$:
    During \textit{Phase 4}, this algorithm is executed by the \textit{Issuer} to issue $\mathtt{VC}$ upon successful verification of the hash ($R$) and proofs ($\boldsymbol{\pi}_{\mathtt{VC}}$, $\boldsymbol{\pi}_{\Phi}$). 
    The algorithm outputs qualification proof ($\boldsymbol{\pi}_{S}$) of $\mathtt{VC}$ and the corresponding DID Token hash ($\mathtt{h}(\mathtt{DIDTok})$).

    \item $\{0,1\}$ $\gets$ $\mathsf{VerifyVC}_{\mathtt{pp}}(\mathtt{DID^H},$$\mathtt{h}(\mathtt{DIDTok}),$$\mathtt{Tr_{IS}},$$ \boldsymbol{\pi}_{S},$ $r^0_I,$ $r_I,$ $\mathtt{Tr_{VC}}')$:
    During \textit{Phase 5}, the algorithm is executed by the \textit{Verifier} to check the current session binding and verify the ZKP ($\boldsymbol{\pi}_{S}$).  
\end{itemize}

\subsection{Public Parameters for Signature-less VC}

In \textit{Phase 1}, the blockchain in SLVC-DIDA generates three public parameters ($\mathtt{pp}_{\Phi}$, $\mathtt{pp}_{S}$, $\mathtt{pp}_{\mathtt{h}}$) for issuance ZKP ($\boldsymbol{\pi}_{\Phi}$), qualification ZKP ($\boldsymbol{\pi}_{S}$), and hash function computation respectively. 
For the issuance ZKP ($\boldsymbol{\pi}_{\Phi}$), the statement $\mathcal{T}$ to be proved by the \textit{Holder} is defined as follows:

\begin{algorithm}[!t]
\caption{$\mathsf{Setup}(\cdot)$ for Initialization}
\begin{algorithmic}[1]
\ENSURE{$\mathtt{pp}$}
\REQUIRE{$\lambda$}

\STATE Decide on issuance predicate $(\Phi_{\mathsf{iss}})$ for VC
\STATE Let $\mathsf{desc}_{\mathcal{T}}$ be a ZKP circuit of statement $\mathcal{T}$ asserting:\\
$\mathtt{VC}$ opens to $(\mathtt{DID^{H}},\mathtt{attr}) $ $\land $ $\Phi_{\mathsf{iss}}(\mathtt{attr}, \mathtt{aux_{ZD}}) = 1 $ $\land $\\ $ \mathtt{attr}$ match the attributes in $w_{\mathtt{ZD}}$

\STATE Let $\mathsf{desc}_{\mathcal{S}}$ be a ZKP circuit of statement $\mathcal{S}$ asserting:\\
$\mathtt{h}(\mathtt{DIDTok}) = \mathsf{CRH.Hash}_{\mathtt{pp_{h}}}$$(\mathtt{DID^{I}},$ $\mathtt{VC},$ $\mathtt{h}(r_H \parallel r_I))$ $ \land $\\
$\mathtt{VC} \in \mathtt{TR_{VC}}'$ $ \land $ $\mathtt{DID^I} \in \mathtt{IS}$  

\STATE Compute $\mathtt{pp}_{\Phi} := \mathsf{NIZK.Setup}(1^\lambda, \mathsf{desc}_{\mathcal{T}})$
\STATE Compute $\mathtt{pp}_S := \mathsf{NIZK.Setup}(1^\lambda, \mathsf{desc}_{\mathcal{S}})$
\STATE Compute $\mathtt{pp_{\mathtt{h}}} := \mathsf{CRH.Setup}(1^\lambda)$
\STATE Construct $\mathtt{pp} := (\mathtt{pp}_{\Phi}, \mathtt{pp}_S,\mathtt{pp_{\mathtt{h}}})$.
\STATE Return $\mathtt{pp}$
\end{algorithmic}
\label{alg:setup}
\end{algorithm}

\begin{mdframed}
\textbf{\underline{Statement $\mathcal{T}$:}}

I know private $\mathtt{attr}$ and $w_{\mathtt{ZD}}$ such that:

\begin{algorithmic}[1]
    \STATE $\mathtt{VC}$ opens to ($\mathtt{attr}$, $\mathtt{DID^{H}}$).
    \STATE $\Phi_{\mathsf{iss}}(\mathtt{attr,\mathtt{aux_{ZD}}})=1$. 
    \STATE $\mathtt{attr}$ match the attributes in $w_{\mathtt{ZD}}$. 
\end{algorithmic}
\end{mdframed}
In the statement $\mathcal{T}$ of $\boldsymbol{\pi}_{\Phi}$, the attributes ($\mathtt{attr}$) and ZKSD witness ($w_{\mathtt{ZD}}$) serve as private witnesses, while the identifier ($\mathtt{DID^{H}}$) and auxiliary metadata ($\mathtt{aux}_{\mathtt{ZD}}$) function as public instances.
The statement $\mathcal{T}$ ensures that the plaintext attributes remain confidential from both \textit{Issuer} and \textit{Verifier}, while guaranteeing the correctness and authority of VCs through through ZKSD ($\mathtt{ZD}$). 
By employing ZKP, our signature-less VC achieves privacy preservation while effectively preventing the generation of falsified/misleading VCs.

Moreover, for the \textit{Issuer} ($\mathtt{DID^I}$) in the legitimate \textit{Issuer} set ($\mathtt{IS}$), it first computes the corresponding Merkle-tree membership proof ($\boldsymbol{\pi}_{\mathtt{IS}}$) based on Merkle-tree public parameters ($\mathtt{pp_{mk}}$).
for the qualification ZKP ($\boldsymbol{\pi}_{S}$), the statement $\mathcal{S}$ of the \textit{Issuer} is defined as follows:

\begin{mdframed}
\textbf{\underline{Statement $\mathcal{S}$:}}

I know private $\mathtt{DID^I}$ (with $\boldsymbol{\pi}_{\mathtt{IS}}$), $\mathtt{VC}$ (with $\boldsymbol{\pi}_{\mathtt{VC}}$) and $r_H$ such that:

\begin{algorithmic}[1]
    \STATE $\mathtt{VC}$ $\in$ $\mathtt{TR_{VC}}'$.
    \STATE 
    $\mathtt{DID^I}$ $\in$ $ \mathtt{TR_{IS}}$.
    \STATE $\mathtt{h}(\mathtt{DIDTok}) = \mathsf{CRH.Hash}_{\mathtt{pp_{h}}}(\mathtt{DID^{I}},\mathtt{VC},\mathtt{h}(r_H \parallel r_I))$.
\end{algorithmic}
\end{mdframed}
On one hand, in the statement $\mathcal{S}$, the \textit{Issuer} identifier ($\mathtt{DID^I}$), the \textit{Issuer} membership proof ($\boldsymbol{\pi}_{\mathtt{IS}}$), the signature-less VC ($\mathtt{VC}$), the VC membership proof ($\boldsymbol{\pi}_{\mathtt{VC}}$) and randomness ($r_H$) serve as private witness. 
On the other hand, the hash value of DID Token ($\mathtt{h}(\mathtt{DIDTok})$), the updated Merkle root ($\mathtt{Tr_{VC}}'$) of VC list, the Merkle root ($\mathtt{Tr_{IS}}$) of \textit{Issuer} set and randomness ($r_I$) are public instances.
The statement $\mathcal{S}$ guarantees the \textit{Issuer} anonymity and verifiability, so that it effectively conceals potentially sensitive information, including any links between signature-less VCs and specific \textit{Issuer} identities. 

Alg. \ref{alg:setup} shows the details of $\mathsf{Setup(\cdot)}$ in \textit{Phase 1}. 
The $\mathsf{Setup(\cdot)}$ algorithm of signature-less VC inputs the security parameter ($\lambda$) and generates the public parameters ($\mathtt{pp}_{\Phi}$, $\mathtt{pp}_{S}$) for the ZKP circuits of statements $\mathcal{T}$ and $\mathcal{S}$, respectively.
All public parameters ($\mathtt{pp}$) are send to the \textit{Holder}, the \textit{Issuer} and the \textit{Verifier} to execute operations of signature-less VC.
In addition, the public parameters ($\mathtt{pp_{mk}}$) of Merkle tree is generated before DID authentication process.

\subsection{Membership proof and Issuance ZKP for Signature-less VC}\label{sec:zero}

In \textit{Phase 3}, the Holder creates the signature-less VC ($\mathtt{VC}$) and the corresponding membership proof ($\boldsymbol{\pi}_{\mathtt{VC}}$) through the $\mathsf{GenVC}_{\mathtt{pp}}(\cdot)$ algorithm.
Before the VC and its proof generation, the \textit{Holder} retrieves the current Merkle tree ($\mathtt{TR_{VC}}$) of VC list, along with its public parameters ($\mathtt{pp_{mk}}$), from the blockchain.
Alg. \ref{alg:genvc} shows the details of the $\mathsf{GenVC}_{\mathtt{pp}}(\cdot)$ algorithm for signature-less VC generation.
Specifically, the algorithm takes as input the attributes ($\mathtt{attr}$), identifier ($\mathtt{DID^{H}}$) and Merkle tree ($\mathtt{TR_{VC}}$). 
First, it samples the commitment nonce ($n$) to create $\mathtt{VC}$, constructing a VC witness ($w_{\mathtt{VC}}$) from ($n$, $\mathtt{attr}$, $\mathtt{DID^{H}}$). 
Subsequently, it updates the on-chain Merkle tree by adding the newly generated $\mathtt{VC}$, thereby computing a modified the Merkle tree ($\mathtt{TR_{VC}}'$) with a new root ($\mathtt{Tr_{VC}}'$). 
Finally, the algorithm computes the membership proof ($\boldsymbol{\pi}_{\mathtt{VC}}$) to prove $\mathtt{VC}$ $\in$ $\mathtt{TR_{VC}}'$, and then outputs both $\mathtt{VC}$ and $\boldsymbol{\pi}_{\mathtt{VC}}$ to the \textit{Holder}.

\begin{algorithm}[!t]
\caption{$\mathsf{GenVC}_{\mathtt{pp}}(\cdot)$ for Signature-less VC Generation}
\begin{algorithmic}[1]
\ENSURE{$\mathtt{VC}$ and $\boldsymbol{\pi}_{\mathtt{VC}}$}

\REQUIRE{$\mathtt{attr}$, $\mathtt{DID^{H}}$ and $\mathtt{TR_{VC}}$ with $\mathtt{pp_{mk}}$}

\STATE Sample nonce $n$ for commitment
\STATE Commit $\mathtt{VC} :=\mathsf{Commit}(\mathtt{attr}, \mathtt{DID^{H}}; n)$
\STATE Construct $w_{\mathtt{VC}} := (n, \mathtt{DID^{H}}, \mathtt{attr})$ for VC witness
\STATE Add new $\mathtt{VC}$ into $\mathtt{TR_{VC}}$ and update the Merkle tree of VC list:
$\mathtt{TR_{VC}}'$ $\gets$ $\mathsf{MK.Add}_{\mathtt{pp_{mk}}}(\mathtt{VC}, \mathtt{TR_{VC}})$
\STATE Compute the authentication path as the membership proof: $\boldsymbol{\pi}_{\mathtt{VC}}$$\gets$$\mathsf{MK.ComMem}_{\mathtt{pp_{mk}}}(\mathtt{VC},\mathtt{TR_{VC}}')$ 

\STATE Return $w_{\mathtt{VC}}$ and ($\mathtt{VC}$, $\boldsymbol{\pi}_{\mathtt{VC}}$)

\end{algorithmic}
\label{alg:genvc}
\end{algorithm}

\begin{algorithm}[!t]
\caption{$\mathsf{AuthReq}_{\mathtt{pp}}(\cdot)$ for Signature-less VC Generation}
\begin{algorithmic}[1]

\ENSURE {$\boldsymbol{\pi}_{\Phi}$}
\REQUIRE {$\mathtt{VC}$, $w_{\mathtt{VC}}$, $\mathtt{aux_{ZD}}$ and $w_{\mathtt{ZD}}$}

\STATE Construct $w_{\Phi}:=(w_{\mathtt{VC}},w_{\mathtt{ZD}})$ for issuance witness

\STATE Based on $\mathtt{pp}_{\Phi}$ of statement $\mathcal{T}$, compute the issuance ZKP of $\mathtt{VC}$:
$\boldsymbol{\pi}_{\Phi} \gets \mathsf{NIZK.Prove}_{\mathtt{pp}_{\Phi}}((\mathtt{VC}, \mathtt{aux_{ZD}}), w_{\Phi})$

\STATE Return $\boldsymbol{\pi}_{\Phi}$

\end{algorithmic}
\label{alg:authreq}
\end{algorithm}

\begin{algorithm}[!t]
\caption{$\mathsf{VerifyIss}_{\mathtt{pp}}(\cdot)$ for Signature-less VC Issuance}
\begin{algorithmic}[1]
\ENSURE a bit value of $0/1$
\REQUIRE { $\mathtt{VC}$, $\boldsymbol{\pi}_{\mathtt{VC}}$, $\boldsymbol{\pi}_{\Phi}$, $\mathtt{Tr_{VC}}'$ with $\mathtt{pp_{mk}}$, $\mathtt{aux_{ZD}}$, $R$, $r_H$, $r_I$}

\STATE Based on $\mathtt{pp_{h}}$, compute the hash value of randomness:\\
$\mathtt{h}(r_H \parallel r_I)$ $\gets$ $\mathsf{CRH.Hash}_{\mathtt{pp_{h}}}(r_H \parallel r_I)$

\STATE Check $R$ $\overset{?}{=}$ $\mathtt{h}(r_H \parallel r_I)$ for session binding

\STATE Based on $\mathtt{pp_{mk}}$ of tree $\mathtt{TR_{VC}}'$, verify the membership proof:\\
$\{ 0,1\}$ $\gets$ $\mathsf{MK.Verify}_{\mathtt{pp_{mk}}}(\mathtt{VC}, \boldsymbol{\pi}_{\mathtt{VC}}, \mathtt{Tr_{VC}}')$

\STATE Based on $\mathtt{pp}_{\Phi}$ of statement $\mathcal{T}$, verify the issuance ZKP:\\
$\{ 0,1\}$ $\gets$ $\mathsf{NIZK.Verify}_{\mathtt{pp}_{\Phi}}( \boldsymbol{\pi}_{\Phi}, (\mathtt{VC}, \mathtt{aux_{ZD}}))$

\STATE If either verification fails, stop the VC issuance and reject the authentication request

\STATE Return $1$ for successful verification; otherwise, return $0$

\end{algorithmic}
\label{alg:veriss}
\end{algorithm}

\begin{algorithm}[!t]
\caption{$\mathsf{IssueVC}_{\mathtt{pp}}(\cdot)$ for Signature-less VC Issuance}
\begin{algorithmic}[1]
\ENSURE $\boldsymbol{\pi}_{S}$ and $\mathtt{h}(\mathtt{DIDTok})$
\REQUIRE{$\mathtt{VC}$, $\mathtt{DID^{I}}$, $\mathtt{TR_{IS}}$, $\mathtt{pp_{mk}}$, $\mathtt{Tr_{IS}}$, $\boldsymbol{\pi}_{\mathtt{VC}}$, $\mathtt{Tr_{VC}}'$, $R$, $r_H$, $r_I$}

\STATE Construct $\mathtt{DIDTok}:= (\mathtt{DID^{I}},\mathtt{VC},R)$

\STATE Based on $\mathtt{pp_{h}}$, compute the hash value of DID Token:
$\mathtt{h}(\mathtt{DIDTok}) \gets \mathsf{CRH.Hash}_{\mathtt{pp_{h}}}(\mathtt{DID^I}, \mathtt{VC}, \mathtt{h}(r_H \parallel r_I))$

\STATE Based on $\mathtt{pp_{mk}}$, compute the membership proof of $\mathtt{DID^I}$:
$\boldsymbol{\pi}_{\mathtt{IS}}$ $\gets$ $\mathsf{MK.ComMem}_{\mathtt{pp_{mk}}}(\mathtt{DID^{I}},\mathtt{TR_{IS}})$

\STATE Construct $w_{S}$ $:=$ $(r_{H}, R, \mathtt{VC}, \boldsymbol{\pi}_{\mathtt{VC}}, \mathtt{DID^{I}}, \boldsymbol{\pi}_{\mathtt{IS}})$ for quantification witness

\STATE Construct $\mathtt{aux}_{S}$ $:=$ $(r_{I}, \mathtt{h}(\mathtt{DIDTok}), \mathtt{Tr_{VC}}', \mathtt{Tr_{IS}})$ for quantification instance

\STATE Based on $\mathtt{pp}_{S}$ of statement $\mathcal{S}$, compute the qualification ZKP of $\mathtt{VC}$:
$\boldsymbol{\pi}_{S} \gets \mathsf{NIZK.Prove}_{\mathtt{pp}_{S}}(\mathtt{aux}_{S}, w_{S})$

\STATE Return $\boldsymbol{\pi}_{S}$ and $\mathtt{h}(\mathtt{DIDTok})$

\end{algorithmic}
\label{alg:issvc}
\end{algorithm}

To ensure the correctness and authority of the signature-less VC generated by the \textit{Holder}, the \textit{Holder} generates an issuance ZKP ($\boldsymbol{\pi}_{\Phi}$) based on ZKSD ($\mathtt{ZD}$) to prove that the $\mathtt{VC}$ is derived from an authoritative credential and satisfies the issuance predicate ($\Phi_{\mathsf{iss}}$).
Such ZKP ($\boldsymbol{\pi}_{\Phi}$) enables the \textit{Issuer} to verify the validation of $\mathtt{VC}$ without compromising privacy, thereby facilitating the issuance of the $\mathtt{VC}$.
Alg. \ref{alg:authreq} specifies the $\mathsf{AuthReq}_{\mathtt{pp}}(\cdot)$ process for creating the issuance ZKP ($\boldsymbol{\pi}_{\Phi}$) of $\mathtt{VC}$.
The algorithm first constructs the issuance witness ($w_{\Phi}$) by combining the VC witness ($w_{\mathtt{VC}}$) and the ZKSD witness ($w_{\mathtt{ZD}}$).
Meanwhile, the tuple ($\mathtt{VC}$, $\mathtt{aux_{ZD}}$) serves as public instance of issuance ZKP ($\boldsymbol{\pi}_{\Phi}$). 
Based on the public parameters ($\mathtt{pp}_{\Phi}$) of the statement $\mathcal{T}$, the algorithm outputs the issuance ZKP ($\boldsymbol{\pi}_{\Phi}$) of the signature-less VC.
After generating both proofs ($\boldsymbol{\pi}_{\mathtt{VC}}$, $\boldsymbol{\pi}_{\Phi}$), the \textit{Holder} sends the signature-less VC ($\mathtt{VC}$ with $\boldsymbol{\pi}_{\Phi}$ and $\boldsymbol{\pi}_{\mathtt{VC}}$) and auxiliary information ($r_H$, $R$, $\mathtt{aux_{ZD}}$) to the \textit{Issuer}.

\subsection{Qualification ZKP for Signature-less VC}


In \textit{Phase 4}, the \textit{Issuer} first retrieves the public Merkle root ($\mathtt{Tr_{VC}}'$) along with the public parameters ($\mathtt{pp_{mk}}$) from the blockchain. 
It then verifies the proofs ($\boldsymbol{\pi}_{\Phi}$ and $\boldsymbol{\pi}_{\mathtt{VC}}$) for the $\mathtt{VC}$ by invoking the $\mathsf{VerifyIss}_{\mathtt{pp}}(\cdot)$ algorithm. 
As detailed in Alg. \ref{alg:veriss}, the $\mathsf{VerifyIss}_{\mathtt{pp}}(\cdot)$ algorithm enables the \textit{Issuer} to validate the correctness and authority of the signature-less VC. 
Based on the public parameters ($\mathtt{pp_h}$), the algorithm first checks whether the randomness ($R$) matches the hash of the concatenated randomness ($r_H \parallel r_I$). 
Such check ensures that the $\mathtt{VC}$ is bound to the current session to prevent replay attacks.  
Subsequently, the algorithm executes $\mathsf{MK.Verify}(\cdot)$ to verify the on-chain membership of $\mathtt{VC}$ in the updated tree ($\mathtt{TR_{VC}}'$).
Additionally, it verifies the issuance proof ($\boldsymbol{\pi}_{\Phi}$) through the $\mathsf{NIZK.Verify}(\cdot)$ with public parameters ($\mathtt{pp}_{\Phi}$) to confirm that the statement $\mathcal{T}$ holds. 
If any verification fails, it rejects the \textit{Holder}'s authentication request and returns $0$; returns $1$ upon full success.

\begin{figure}[!t]
 \centerline{\includegraphics[width=1.0\linewidth]{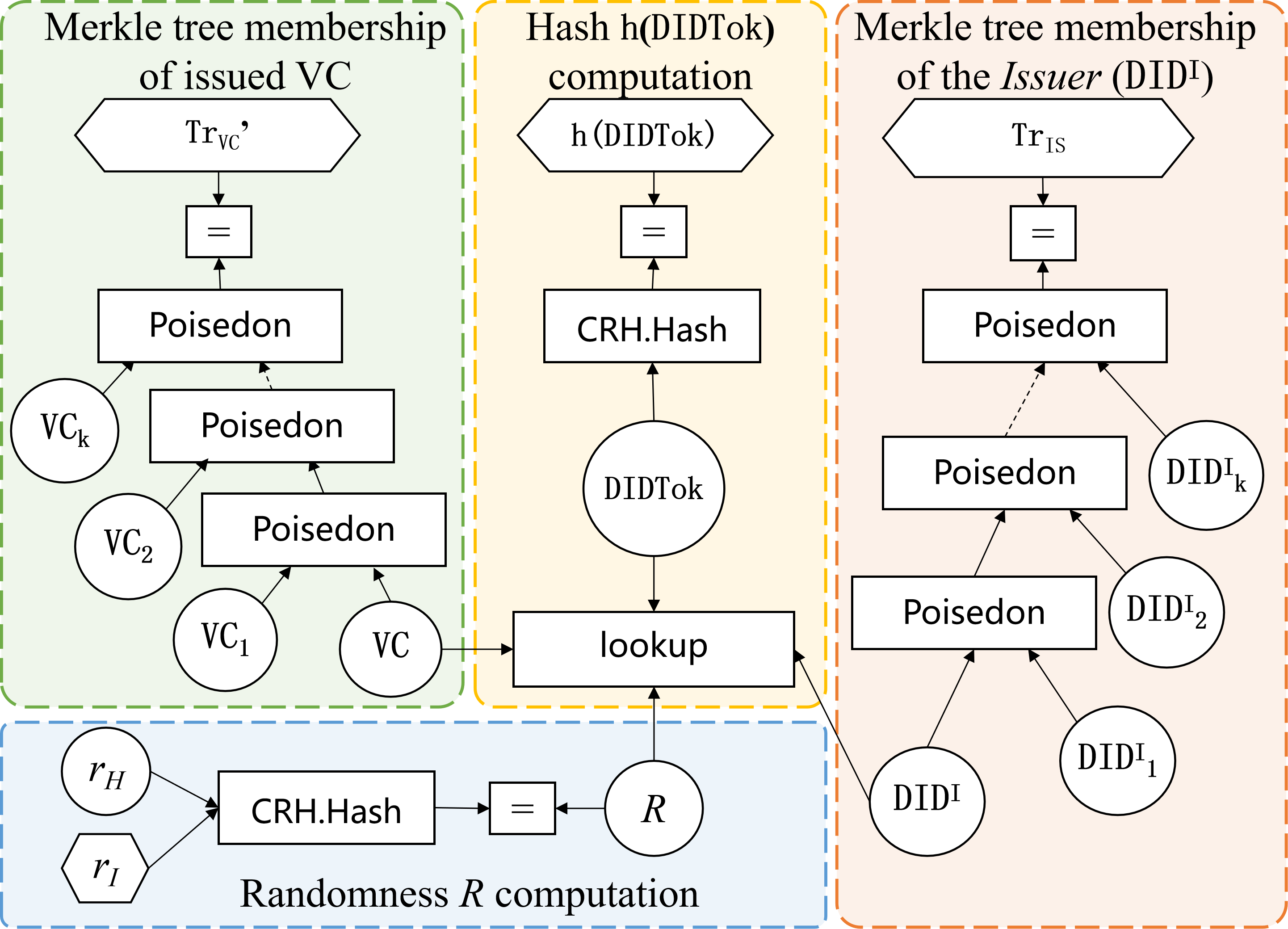}}
 \caption{ZKP circuit of statement $\mathcal{S}$. 
Circular nodes represent private inputs, while hexagonal nodes represent public inputs. Our CRH function, Merkle trees are realized by Poisedon hash in our SLVC-DIDA. Each arrow represents a group of wires.} \label{fig:ZK}
\end{figure}

Upon successful verification in Alg. \ref{alg:veriss}, the \textit{Issuer} creates the qualification ZKP ($\boldsymbol{\pi}_{S}$) through the $\mathsf{IssueVC}_{\mathtt{pp}}(\cdot)$ algorithm.
Alg. \ref{alg:issvc} shows the details of the $\mathsf{IssueVC}_{\mathtt{pp}}(\cdot)$ algorithm for the signature-less VC issuance.
Specifically, the signature-less VC issuance process constructs three parameters, i.e., DID Token ($\mathtt{DIDTok}$), quantification witness ($w_S$) and quantification instance ($\mathtt{aux}_S$).
It generates the quantification ZKP ($\boldsymbol{\pi}_{S}$) based on the public parameters $\mathtt{pp}_{S}$, and outputs the ZKP ($\boldsymbol{\pi}_{S}$) of $\mathtt{VC}$ to demonstrate the validity of statement $\mathcal{S}$.

As shown in Fig. \ref{fig:ZK}, the ZKP circuit for qualification proof ($\boldsymbol{\pi}_{S}$) fully implements the verification process of statement $\mathcal{S}$.
This ZKP circuit not only validates the signature-less VC and verifies the \textit{Issuer}'s legitimacy, but also ensures issuer-hiding by privately linking the \textit{Issuer}'s identity to the issued $\mathtt{VC}$; thus, the \textit{Issuer}'s anonymity within the DID authentication process is preserved. 
We instantiate the algorithm \( \mathsf{CRH.Hash}_{\mathtt{pp}}(\cdot) \) by using the Poseidon hash function and compute the randomness (\( R \)) through the public randomness input (\( r_{I} \)) and the private randomness input (\( r_{H} \)). 
Computed from the random sampling, randomness (\( r_{I} \), \( r_{H} \)) are used to enhance the unpredictability of cryptographic operations. 
Furthermore, the $\mathsf{lookup}(\cdot)$ function verifies whether the identity token ($\mathtt{DIDTok}$) contains the correct $R$, $\mathtt{DID}^{I}$, and $\mathtt{VC}$. 
The identity token ($\mathtt{DIDTok}$) ensures the privacy and integrity of the VC assertions by disclosing only the hash value ($\mathtt{h}(\mathtt{DIDTok})$).
For the Merkle tree membership of the issued VC ($\mathtt{VC}$), the authentication path from the $\mathtt{VC}$ to the Merkle root ($\mathtt{Tr_{VC}}'$) is used as VC membership proof ($\boldsymbol{\pi}_{\mathtt{VC}}$).
In particular, the ZKP circuit configures the sibling nodes ($\mathtt{VC}_1$, $\mathtt{VC}_2$,...,$\mathtt{VC}_k$) (where $k$ denotes the Merkle tree depth) as private inputs, while the Merkle root ($\mathtt{Tr_{VC}}'$) remains a public input. 
It enables formal verification of $\mathtt{VC} \in \mathtt{TR_{VC}}'$ through disclosure of solely the root value ($\mathtt{Tr_{VC}}'$), achieving both proof completeness and privacy preservation.
The final part of the statement $\mathcal{S}$ is the Merkle-tree membership of the legitimate \textit{Issuer} (\( \mathtt{DID}^{I} \)) with \( \mathtt{DID}^{I} \) $\in$ \( \mathtt{IS} \). 
The Merkle-tree membership proof ($\boldsymbol{\pi}_{\mathtt{IS}}$) of the \textit{Issuer} ($\mathtt{DID}^{I}$) consists of the identifiers of the sibling nodes along the authentication path (\( \mathtt{DID}_{1}^{I}, \mathtt{DID}_{2}^{I}, \dots, \mathtt{DID}_{k}^{I} \)). 
This proof ($\boldsymbol{\pi}_{\mathtt{IS}}$) is provided as a private witness to the ZKP circuit, asserting that the qualification of Issuer ($\mathtt{DID}^{I}$) is legitimate without revealing any auxiliary information beyond the public Merkle root (\( \mathtt{Tr_{IS}} \)).



Finally, the \textit{Issuer} sends the ZKP ($\boldsymbol{\pi}_{S}$), the hash value ($\mathtt{h}(\mathtt{DIDTok})$) and the randomness ($r_I^0$) to the \textit{Holder}.
The \textit{Holder} then stores the tuple ($\boldsymbol{\pi}_{S}$, $\mathtt{h}(\mathtt{DIDTok})$, $r_I^0$, $r_I$) on the blockchain (excluding the tree $\mathtt{TR_{VC}}'$ and $\mathtt{TR_{IS}}$) for subsequent verification process.

\subsection{ZKP Verification for Signature-less VC}

In \textit{Phase 5}, the \textit{Verifier} verifies the qualification ZKP ($\boldsymbol{\pi}_{S}$) upon receiving the verification request from the \textit{Holder} ($\mathtt{DID^{H}}$).
The \textit{Verifier} retrieves the corresponding tuple ($\boldsymbol{\pi}_{S}$, $\mathtt{h}(\mathtt{DIDTok})$, $r_I^0$, $r_I$) containing the ZKP ($\boldsymbol{\pi}_{S}$) from the blockchain. 
Moreover, Alg. \ref{alg:verify} executes $\mathsf{VerifyVC}_{\mathtt{pp}}(\cdot)$ to verify the signature-less VC. 
Specifically, it takes as inputs \textit{Holder} identifier ($\mathtt{DID^H}$), token hash ($\mathtt{h}(\mathtt{DIDTok})$), Merkle root ($\mathtt{Tr_{IS}}$) of \textit{Issuer} set, qualification ZKP ($\boldsymbol{\pi}_{S}$), randomness ($r^0_I$, $r_I$), and Merkle root ($\mathtt{Tr_{VC}}'$) of VC list.
It first computes the hash ($\mathtt{h}(\mathtt{DID^H}, r_I^0)$) using public parameters ($\mathtt{pp_{h}}$), then verifies session binding via $r_I \overset{?}{=} \mathsf{CRH.Hash}_{\mathtt{pp_{h}}}(\mathtt{DID^H}, r_I^0)$. 
Subsequently, it parses the quantification instance $\mathtt{aux}_{S} := (r_{I}, \mathtt{h}(\mathtt{DIDTok}), \mathtt{Tr_{VC}}', \mathtt{Tr_{IS}})$ as public inputs and verifies qualification ZKP ($\boldsymbol{\pi}_{S}$) for statement $\mathcal{S}$ using the $\mathsf{NIZK.Verify}_{\mathtt{pp}_{S}}(\cdot)$ algorithm.
Rejection occurs immediately if either session binding or ZKP verification fails. 
Successful verification returns $1$ only upon passing all checks, otherwise returning $0$.

\begin{algorithm}[!t]
\caption{$\mathsf{VerifyVC}_{\mathtt{pp}}(\cdot)$ for Signature-less VC Verification}
\begin{algorithmic}[1]
\ENSURE a bit value of $0/1$
\REQUIRE $\mathtt{DID^H}$, $\mathtt{h}(\mathtt{DIDTok})$, $\mathtt{Tr_{IS}}$, $ \boldsymbol{\pi}_{S}$, $r^0_I$, $r_I$, $\mathtt{Tr_{VC}}'$

\STATE Based on $\mathtt{pp_{h}}$, compute the hash value of randomness and identifier:
$\mathtt{h}(\mathtt{DID^H}, r_I^0)$ $\gets$ $\mathsf{CRH.Hash}_{\mathtt{pp_{h}}}(\mathtt{DID^H}, r_I^0)$

\STATE Check $r_I$ $\overset{?}{=}$ $\mathsf{CRH.Hash}_{\mathtt{pp_{h}}}(\mathtt{DID^H}, r_I^0)$ for session binding

\STATE Parse $(r_{I}, \mathtt{h}(\mathtt{DIDTok}), \mathtt{Tr_{VC}}', \mathtt{Tr_{IS}})$ $:=$ $\mathtt{aux}_{S}$

\STATE Based on $\mathtt{pp}_{S}$ of statement $\mathcal{S}$, verify qualification ZKP:
$\{0,1\}$ $\gets$ $\mathsf{NIZK.Verify}_{\mathtt{pp}_{S}}(\boldsymbol{\pi}_{S},\mathtt{aux}_{S})$

\STATE If either verification fails, reject the verification request

\STATE Return $1$ for successful verification; otherwise, return $0$
\end{algorithmic}
\label{alg:verify}
\end{algorithm}

Since the qualification proof ($\boldsymbol{\pi}_{S}$) asserts the validity of the VC and the legitimacy of the \textit{Issuer}, our signature-less VC enables zero-knowledge verification where the \textit{Verifier} can validate VCs without accessing their content, while commitments ensure the confidentiality of the \textit{Holder}'s attributes. 
Moreover, our signature-less VC realizes a privacy-preserving and issuer-hiding DID authentication that fully preserves VC privacy while completely concealing \textit{Issuer} identities.
\section{Security Analysis}


We define Unforgeability, Issuance Privacy, Verification Privacy and Issuer Anonymity in this section and prove that our signature-less VC scheme holds these properties.
Specifically, we allow an adversary's possible access to a generation oracle ($\mathcal{GO}$) and a issuance oracle ($\mathcal{IO}$), defined as follows:

\begin{definition}[Generation Oracle ($\mathcal{GO}$)]
Initialize the set of generated VCs by $\boldsymbol{\Pi}_{\mathcal{GO}} = \emptyset$ and the Merkle tree ($\mathtt{TR_{VC}}$) via $\mathsf{MK.TreeGen}_{\mathtt{pp_{mk}}}(\emptyset)$. 
When $\mathcal{GO}$ is queried with attributes ($\mathtt{attr}$), an identifier ($\mathtt{DID^H}$), public auxiliary metadata ($\mathtt{aux}_{\mathtt{pub}}$) including $\mathtt{aux_{ZD}}$ and $r_I$, and private auxiliary metadata ($\mathtt{aux}_{\mathtt{pri}}$) including $w_\mathtt{{ZD}},$ $r_H$ and $R$, $\mathcal{GO}$ returns a valid VC generation tuple $(\mathtt{VC}, \boldsymbol{\pi}_{\mathtt{VC}}, \boldsymbol{\pi}_{\Phi})$ and adds it to $\boldsymbol{\Pi}_{\mathcal{GO}}$, where VC and membership proof are generated via $(\mathtt{VC},\boldsymbol{\pi}_{\mathtt{VC}}) \leftarrow $ $\mathsf{GenVC}_{\mathtt{pp}}(\mathtt{attr},\mathtt{DID^{H}},\mathtt{TR_{VC}};\mathtt{pp_{mk}})$ (which updates $\mathtt{TR_{VC}}$ and its root $\mathtt{Tr'_{VC}}$), and the issuance proof is generated via $\boldsymbol{\pi}_{\Phi} \leftarrow $ $\mathsf{AuthReq}_{\mathtt{pp}}(\mathtt{VC},\mathtt{aux}_{\mathtt{pub}},\mathtt{aux}_{\mathtt{pri}})$.

\end{definition}

\begin{definition}[Issuance Oracle ($\mathcal{IO}$)]
Initialize a set of issued $\boldsymbol{\Pi}_{\mathcal{IO}} = \emptyset$. The oracle $\mathcal{IO}$ shares the queries ($\boldsymbol{\Pi}_{\mathcal{GO}}$) with generation oracle ($\mathcal{GO}$).
Upon receiving a query with inputs $(\mathtt{VC}, \mathtt{DID^{I}},$ $ \mathtt{IS},$ $ \mathtt{Tr_{IS}}, \boldsymbol{\pi}_{\mathtt{VC}},$ $ \mathtt{Tr_{VC}}', $ $R, r_H, r_I, $ $\mathtt{pp_{mk}}, $ $\boldsymbol{\pi}_{\Phi})$, $\mathcal{IO}$ first validates authorization by checking if $(\mathtt{VC}, \boldsymbol{\pi}_{\mathtt{VC}},$ $\boldsymbol{\pi}_{\Phi}) $ $\in \boldsymbol{\Pi}_{\mathcal{GO}}$; if the check fails, it returns $\bot$. 
Otherwise, it executes $(\boldsymbol{\pi}_S, $ $\mathtt{h}(\mathtt{DIDTok})) \leftarrow $ $\mathsf{IssueVC}_{\mathtt{pp}}(\mathtt{VC}, $ $\mathtt{DID^{I}}, \mathtt{IS},$ $ \mathtt{Tr_{IS}}, \boldsymbol{\pi}_{\mathtt{VC}},$ $ \mathtt{Tr_{VC}}', R, r_H, r_I, $ $\mathtt{pp_{mk}})$, and returns the qualification proof ($\boldsymbol{\pi}_{S}$) along with the DID Token hash ($\mathtt{h}(\mathtt{DIDTok})$). Moreover, it updates $\boldsymbol{\Pi}_{\mathcal{IO}}$ by adding $(\boldsymbol{\pi}_{S},$ $\mathtt{h}(\mathtt{DIDTok}))$.

\end{definition}

\begin{definition}[VC Generation Unforgeability]
Our generated signature-less VC in Phase 3 holds unforgeability if for any PPT adversary $\mathcal{A}$ and security parameter $\lambda$, there exists a negligible function $\mathsf{negl}$ such that:
\[
\Pr\left[
\begin{matrix}
    \mathtt{pp} \gets \mathsf{Setup}(1^{\lambda});
    \mathtt{pp_{mk}} \gets \mathsf{MK.Setup}(1^{\lambda});\\
    \mathtt{TR}^*_{\mathtt{VC}} \leftarrow \mathsf{MK.TreeGen}_{\mathtt{pp_{mk}}}(\emptyset);\\
    (\mathtt{VC}^*,\boldsymbol{\pi}_{\mathtt{VC}}^*,\boldsymbol{\pi}_{\Phi}^*,\mathtt{Tr}'^*_{\mathtt{VC}}) \gets \mathcal{A}^{\mathcal{GO}}_{\mathtt{pp}}
    \left(
    \begin{matrix}
        \mathtt{pp_{mk}},\mathtt{TR}^*_{\mathtt{VC}}\\
    \end{matrix}
    \right):
    \\
    \mathsf{VerifyIss}_{\mathtt{pp}} \left(
    \begin{matrix}
    \mathtt{pp_{mk}}, 
    \mathtt{VC}^*,\boldsymbol{\pi}_{\mathtt{VC}}^*, r_I\\
    \boldsymbol{\pi}_{\Phi}^*,\mathtt{aux}^*_{\mathtt{pub}},\mathtt{Tr}'^*_{\mathtt{VC}} 
    \end{matrix}
    \right) = 1 \\
    \land \ (\mathtt{VC}^*,\boldsymbol{\pi}_{\mathtt{VC}}^*,\boldsymbol{\pi}_{\Phi}^*) \notin \boldsymbol{\Pi}_{\mathcal{GO}}
\end{matrix}
\right]
\leq \mathsf{negl}(\lambda).
\]
where $\mathtt{Tr}'^*_{\mathtt{VC}}$ represents the final root of the Merkle tree after all oracle queries.
    
\end{definition}

\begin{definition}[VC Issuance Unforgeability]
Our issued signature-less VC in Phase 4 holds unforgeability if for any PPT adversary $\mathcal{A}$ and security parameter $\lambda$:
\[
\Pr \left[ 
\begin{array}{l}
    \mathtt{pp} \leftarrow \mathsf{Setup}(1^\lambda); \mathtt{pp_{mk}} \gets \mathsf{MK.Setup}(1^{\lambda}); \\
    
    \left(
    \begin{matrix}
        (\boldsymbol{\pi}_S^*, \mathtt{h}(\mathtt{DIDTok})^*), \\
        (\mathtt{VC}^*,\pi_{\mathtt{VC}}^*,\pi_{\Phi}^*), r^{0*}_I, \\
        r^*_I,\mathtt{DID}^*, \mathtt{Tr}^*_{\mathtt{IS}}, \mathtt{Tr}'^*_{\mathtt{VC}}
    \end{matrix}
    \right) 
    \leftarrow \mathcal{A}^{\mathcal{IO},\mathcal{GO}}_\mathtt{pp}(\mathtt{pp_{mk}}): \\
     \mathsf{VerifyVC}_\mathtt{pp}\left(
        \begin{aligned}
            &(\boldsymbol{\pi}_{S}^*, \mathtt{h}(\mathtt{DIDTok})^*), r^{0*}_I,\\
            &\mathtt{Tr}^*_{\mathtt{IS}}, \mathtt{DID}^*,  r^*_I, \mathtt{Tr}'^*_{\mathtt{VC}}
        \end{aligned}
    \right) = 1 \\
     \land \quad (\boldsymbol{\pi}_S^*, \mathtt{h}^*(\mathtt{DIDTok})) \notin \boldsymbol{\Pi}_{\mathcal{IO}} \\
      \land\ \quad \exists\, (\mathtt{VC}, \pi_{\mathtt{VC}}, \pi_{\Phi}) \in \boldsymbol{\Pi}_{\mathcal{GO}}
\end{array} 
\right] \le \mathsf{negl}(\lambda).
\]

\end{definition}

\begin{definition}[VC Issuance Privacy]
For any PPT adversary $\mathcal{A}$ there exists a negligible function $\mathsf{negl}(\cdot)$ such that
\[
\Pr\!\left[
\begin{aligned}
& \mathtt{pp} \leftarrow \mathsf{Setup}(1^{\lambda}); \mathtt{pp_{mk}} \leftarrow \mathsf{MK.Setup}(1^{\lambda});\\
&\mathtt{TR_{VC}} \leftarrow \mathsf{MK.TreeGen}_{pp_{mk}}(\emptyset); b \xleftarrow{\$}\{0,1\};\\
& \left(
\begin{matrix}
    (\mathtt{attr}_0,\mathtt{aux}^{(0)}_{\sf pub},\mathtt{aux}^{(0)}_{\sf pri}), \\
    (\mathtt{attr}_1,\mathtt{aux}^{(1)}_{\sf pub},\mathtt{aux}^{(1)}_{\sf pri}) 
\end{matrix} 
  \right) \leftarrow \mathcal{A}_{\mathtt{pp}}(\mathtt{pp_{mk}});\\
& \left( 
\begin{matrix}
    \mathtt{aux}^{(b)}_{\sf pub}=(\mathtt{aux}^{(b)}_{\sf ZD}, r_I),\\
\mathtt{aux}^{(b)}_{\sf pri}=(w^{(b)}_{\sf ZD}, r_H^{(b)}, R^{(b)}),\\
\Phi_{\sf iss}(\mathtt{attr}_{(b)},\mathtt{aux}^{(b)}_{\sf ZD})=1;
\end{matrix}
\right);\\ 
& 
\left(
\begin{matrix} \mathtt{VC}_b,\boldsymbol{\pi}_{\mathtt{VC},b},\\
\boldsymbol{\pi}_{\Phi,b},\mathtt{Tr'_{VC}}
\end{matrix}
\right) \leftarrow 
   \mathcal{GO}
   \left(
   \begin{matrix}
       \mathtt{attr}_b,\mathtt{DID^H}, \\
       \mathtt{aux}^{(b)}_{\sf pub}, \mathtt{aux}^{(b)}_{\sf pri}
   \end{matrix}
   \right);\\
& V_b \gets \left(
\begin{matrix}
    \mathtt{VC}_b,\boldsymbol{\pi}_{\mathtt{VC},b},\boldsymbol{\pi}_{\Phi,b},\mathtt{Tr'_{VC}},\\
    \mathtt{aux}^{(b)}_{\sf ZD},R^{(b)},r_H^{(b)},r_I
\end{matrix}
\right);\\
& b' \leftarrow \mathcal{A}^{\mathcal{GO},\mathcal{IO}}(V_b):\ b'=b
\end{aligned}
\right] \le \tfrac{1}{2} + \mathsf{negl}(\lambda).
\]
\end{definition}

\begin{definition}[Issuer Anonymity]
For any PPT adversary $\mathcal{A}$ there exists a negligible function $\mathsf{negl}(\cdot)$ such that
\[
\Pr\!\left[
\begin{aligned}
& \mathtt{pp} \leftarrow \mathsf{Setup}(1^{\lambda}); \mathtt{pp_{mk}} \leftarrow \mathsf{MK.Setup}(1^{\lambda});\\
& (\mathtt{VC},\boldsymbol{\pi}_{\mathtt{VC}},\boldsymbol{\pi}_{\Phi}) \in \boldsymbol{\Pi}_{\mathcal{GO}};\ 
  (\mathtt{DID^{I}_0},\mathtt{DID^{I}_1}) \in \mathtt{IS};\\
& \left(
\begin{matrix}
    \boldsymbol{\pi}_{S,b},\\ \mathtt{h(DIDTok)}_b
\end{matrix}
 \right) \leftarrow 
   \mathcal{IO}\left(
   \begin{matrix}
       \mathtt{VC},\mathtt{DID^{I}_b},\mathtt{IS},r_I,\\
       \pi_{\mathtt{VC}},\mathtt{Tr'_{VC}},R,r_H,\\
       \mathtt{Tr_{IS}},\mathtt{pp_{mk}},\boldsymbol{\pi}_{\Phi}
   \end{matrix}
   \right);\\
& b \xleftarrow{\$}\{0,1\}; W_b \gets \left(
\begin{matrix}
    \boldsymbol{\pi}_{S,b}, \mathtt{h(DIDTok)}_b, r_I^0,\\
    r_I, \mathtt{Tr_{IS}}, \mathtt{Tr'_{VC}}, \mathtt{DID^H}
\end{matrix}
\right);\\
& b' \leftarrow \mathcal{A}^{\mathcal{GO},\mathcal{IO}}(W_b):\ b'=b
\end{aligned}
\right] \le \tfrac{1}{2} + \mathsf{negl}(\cdot).
\]
\end{definition}

\begin{definition}[VC Verification Privacy]
Let the public statement be $y=(\mathtt{DID^H},\mathtt{h}(\mathtt{DIDTok}),\mathtt{Tr_{IS}},\mathtt{Tr'_{VC}},r_I)$.
For any PPT adversary $\mathcal{A}$, there exists a PPT simulator $\mathsf{Sim}$ such that:
\[
\begin{aligned}
    & \left|
    \Pr\!\left[
    \begin{aligned}
    & (\pi_S, r_I^0) \leftarrow \mathcal{IO}^{\mathcal{GO}}(\cdot); \\
    & \mathcal{A}(y, \pi_S, r_I^0) = 1
    \end{aligned}
    \right]
    -
    \Pr\!\left[
    \begin{aligned}
    & (\widehat{\pi}_S, \widehat{r}_I^0) \leftarrow \mathsf{Sim}(y); \\
    & \mathcal{A}(y, \widehat{\pi}_S, \widehat{r}_I^0) = 1
    \end{aligned}
     \right]
     \right|  \\
    & \hspace{20em} \le \mathsf{negl}(\cdot) .
\end{aligned}
\]
Here, $\mathcal{IO}^{\mathcal{GO}}(\cdot)$ represents the honest execution of the protocol using the valid VC state obtained from oracles ($\mathcal{IO}$ and $\mathcal{GO}$), and $\mathsf{Sim}$ generates a sproof $(\widehat{\pi}_S, \widehat{r}_I^0)$ using only the public statement $y$, such that $\mathsf{VerifyVC}_{\mathtt{pp}}(y, \widehat{\pi}_S, \widehat{r}_I^0)=1$.

\end{definition}

\begin{theorem}
    Our signature-less VC scheme satisfies VC Generation Unforgeability, VC Issuance Unforgeability, VC Issuance Privacy, VC Verification Privacy, and Issuer Hiding, assuming the collision resistance of CRH, the completeness, soundness, and zero-knowledge of NIZK, the hiding and binding of commitment scheme, and the secure binding of session parameters ($r_I^0$, $r_H$, $R$).
\end{theorem}

\begin{proof} 
    For VC Generation Unforgeability, the proof relies on the collision resistance of CRH and the soundness of the NIZK in $\mathsf{AuthReq}(\cdot)$. 
    We construct a reduction algorithm $B$ that runs the adversary $\mathcal{A}$ in the generation experiment. If $\mathcal{A}$ successfully outputs a tuple $(\mathtt{VC}^*,\boldsymbol{\pi}_{\mathtt{VC}}^*,\boldsymbol{\pi}_{\Phi}^*,\mathtt{Tr}'^*_{\mathtt{VC}})$ that is accepted by the $\mathsf{VerifyIss}(\cdot)$ but was not produced by the generation oracle ($\mathcal{GO}$), one of two cases must occur. 
    First, if $\mathtt{VC}^*$ is not a valid leaf in the tree rooted at $\mathtt{Tr}'^*_{\mathtt{VC}}$, then $\boldsymbol{\pi}_{\mathtt{VC}}^*$ constitutes a forged Merkle membership proof, implying $B$ can extract a hash collision, breaking CRH. 
    Second, if $\mathtt{VC}^*$ is a valid leaf but the proof $\boldsymbol{\pi}_{\Phi}^*$ is accepted without a valid witness, this directly violates the soundness of NIZK in $\mathsf{AuthReq}(\cdot)$. 
    Since the model dictates that only $\mathcal{GO}$ can update the tree root, unauthorized external updates are impossible. Therefore, $\mathcal{A}$'s success probability is bounded by the negligible probability of breaking CRH or the soundness of $\mathsf{AuthReq}(\cdot)$.

    For VC Issuance Unforgeability, it is proven by reducing the $\mathcal{A}$'s advantage to the soundness of NIZK in $\mathsf{IssueVC}(\cdot)$ and the collision resistance of the hash function ($\mathtt{h}(\cdot)$). Let reduction $B'$ run adversary $A$. If $A$ outputs a valid pair $(\boldsymbol{\pi}_S^*, \mathtt{h}(\mathtt{DIDTok})^*)$ accepted by the verification algorithm but not recorded by the issuance oracle ($\mathcal{IO}$), we analyze the possible breaches. 
    If $\boldsymbol{\pi}_S^*$ is accepted without the issuer actually using their witness, $B'$ breaks the soundness of NIZK in $\mathsf{IssueVC}(\cdot)$. Alternatively, if $\mathcal{A}$ derives a token hash $h(DID_{Tok})^*$ that matches an issued DID token without querying $\mathcal{IO}$, $B'$ must have found a collision in $h(\cdot)$ or forged a witness for $\mathsf{IssueVC}(\cdot)$ not permitted by the predicate. Since the oracle semantics ensure $\mathcal{IO}$ records all successful issuances, a simple replay is excluded. 
    Thus, any advantage for $\mathcal{A}$ implies breaking the underlying cryptographic assumptions, which occurs with negligible probability.

    For VC Issuance Privacy, we employ a hybrid argument relying on the hiding property of commitments and the zero-knowledge of $\mathsf{AuthReq}(\cdot)$. In the real experiment (Hybrid 0), $\mathcal{A}$ sees the view for a chosen attribute bit ($b$). In Hybrid 1, we replace the issuance proof ($\boldsymbol{\pi}_{\Phi}$) with a simulator output; indistinguishability holds due to the ZK property. 
    In Hybrid 2, relying on the hiding property of the commitment scheme within the VC, we replace the attribute-dependent commitments in $\mathtt{VC}_b$ with commitments to random values (or the alternative attribute set). Because the leaf ($\mathtt{VC}_b$) is computed over this hiding commitment, the leaf hashes for $b=0$ and $b=1$ are computationally indistinguishable. 
    Consequently, the deterministic Merkle membership proof $\boldsymbol{\pi}_{\mathtt{VC}}$, which depends only on the leaf hash and the public tree structure, follows an indistinguishable distribution. 
    Finally, in Hybrid 3, session nonces $R, r_H$ are replaced by fresh random values. The views in Hybrid 0 and Hybrid 3 are identically distributed, proving that any non-negligible advantage in distinguishing $b$ implies breaking commitment scheme, NIZK, or CRH.

    For VC Verification Privacy, it is proven via a standard simulation argument. Given a public instance $y$, we construct a PPT simulator $\mathsf{Sim}$ that generates a proof pair $(\widehat{\pi}_S, \widehat{r}_I^0)$ accepted by $\mathsf{VerifyVC}(\cdot)$. 
    The existence of $\mathsf{Sim}$ is guaranteed by the ZK property of $\mathsf{IssueVC}(\cdot)$, ensuring the proof reveals nothing but the predicate's validity. 
    Since $r_I^0$ is independent and random, the simulated distribution is computationally indistinguishable from the real execution. 
    Thus, distinguishing real from simulated views would violate the ZK assumption of $\mathsf{IssueVC}(\cdot)$.

    For Issuer Anonymity, we analyze the issuer-challenge game where the adversary selects two issuers ($\mathtt{DID^{I}_0}, \mathtt{DID^{I}_1}) \in \mathtt{IS}$. The proof proceeds via hybrids where we gradually replace the witness used in the $\mathsf{IssueVC}(\cdot)$ from one issuer to the other, while maintaining the predicate ``issued by someone in $\mathtt{IS}$." 
    By the ZK property of NIZK, the proof $\boldsymbol{\pi}_S^*$ distribution is independent of the specific issuer identifier ($\mathtt{DID^{I}}$).
    The token hash $\mathtt{h(DIDTok)}$ is collision resistance, and public roots reveal no index information. 
    Therefore, any advantage in distinguishing issuers implies breaking the ZK property of $\mathsf{IssueVC}(\cdot)$ or collision resistance of CRH.
\end{proof}

\section{Implementation and Evaluation} \label{sec:exp}


\subsection{Experiment Configuration}

The implementation of our proposed SLVC-DIDA framework is developed using Python for system orchestration and the \verb|Circom| language\footnote{The code is available at \url{https://github.com/iden3/circom}.} for constructing arithmetic circuits. We employ the Groth16 proving scheme instantiated on the BN128 elliptic curve to generate and verify ZKPs. For cryptographic hashing, the Poseidon hash function is utilized for both the internal logic of the ZKP circuits and the hash computations within the Merkle tree structure. Specifically, for the VCs generated by the Holder themselves and the membership proof (\(\boldsymbol{\pi}_{\mathtt{VC}}\)), we implemented a zero-knowledge Merkle tree accumulator based on the Poseidon hash.
The experimental environments are running an Intel(R) Xeon(R) Platinum 8352V CPU @ 2.10GHz with 32 cores and 256GB of memory. To ensure the reliability of our performance metrics, we run 5 times for each experiment and record their average.

For the issuance and verification of signature-less VCs across legal \textit{Issuers}, we implement the ZKP circuit illustrated in Figure~\ref{fig:ZK} for the statement \( \mathcal{S} \), which facilitates zero-knowledge verification of \textit{Issuer} qualification. We evaluate the performance of SLVC-DIDA by varying two primary parameters: the number of VCs, denoted as \( N_{\mathtt{VC}} \in \{2^2, 2^3, \dots, 2^9\} \), and the height of the Merkle tree, denoted as \( H_{\mathtt{MK}} \in \{10, 20, 30, 40, 50, 60\} \). Under these configurations, we comprehensively evaluate the computational time costs for all algorithms, the sizes of the proving/verification keys, and the generated proof sizes.
Furthermore, to demonstrate the practicality of SLVC-DIDA within a blockchain environment, we conducted on-chain tests on the Ethereum network. Specifically, we deployed the verifier smart contract (derived from the Circom), and evaluated the gas consumption required for on-chain proof verification. 
These tests aim to confirm that the deployment and execution costs of our scheme remain within acceptable limits for decentralized applications.

\subsection{Experiment Results for SLVC-DIDA}
\subsubsection{Experiments for Signautre-less VC}

\begin{figure*}[!t]
 \centerline{\includegraphics[width=1.0\textwidth]{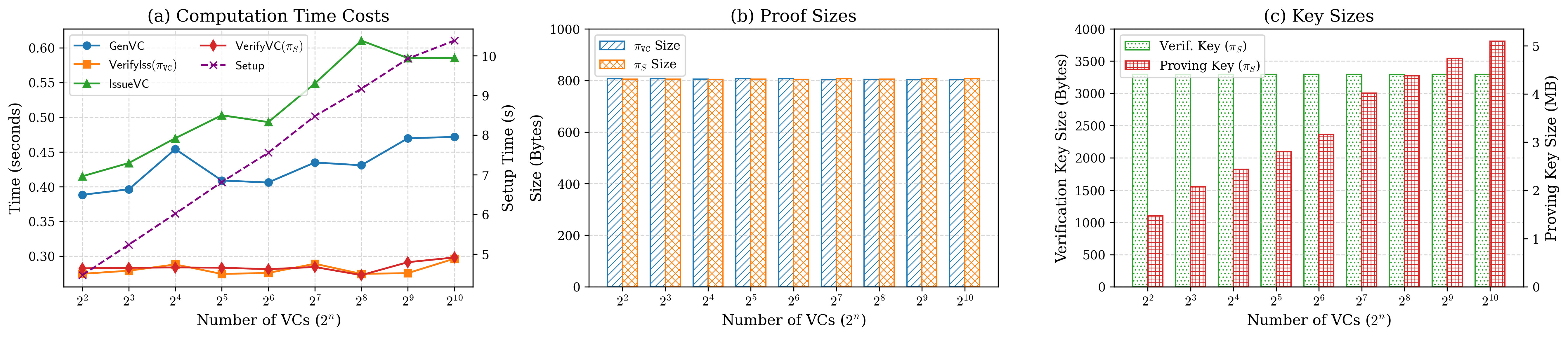}}
 \caption{Performance evaluation of SLVC-DIDA with varying numbers of VCs ($N_{\mathtt{VC}}$). (a) Computation time costs for each algorithm, where $\mathsf{VerifyIss}(\cdot)$ and $\mathsf{VerifyVC}(\cdot)$ measure the verification times for $\boldsymbol{\pi}_{S}$ and $\boldsymbol{\pi}_{\mathtt{VC}}$, respectively; (b) Sizes of the proofs $\boldsymbol{\pi}_{\mathtt{VC}}$ and $\boldsymbol{\pi}_{S}$; (c) Sizes of the proving and verification keys.} \label{fig:ex1}
 \end{figure*}

\begin{figure*}[!t]
 \centerline{\includegraphics[width=1.0\textwidth]{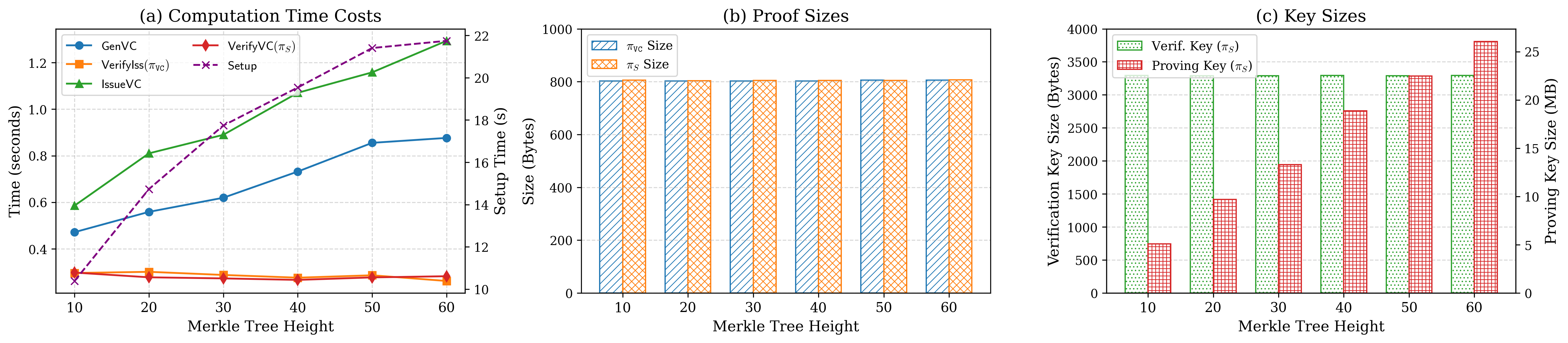}}
 \caption{Performance evaluation of SLVC-DIDA with varying Merkle tree heights ($H_{\mathtt{MK}}$), including (a) time costs of each algorithm, (b) proof sizes and (c) key sizes.} \label{fig:ex2}
 \end{figure*}

Figs. \ref{fig:ex1} and \ref{fig:ex2} illustrate the performance of SLVC-DIDA under varying numbers of VCs ($N_{\mathtt{VC}}$) and Merkle tree heights ($H_{\mathtt{MK}}$), respectively. The results indicate that while the computation costs for $\mathsf{Setup}(\cdot)$, $\mathsf{GenVC}(\cdot)$, and $\mathsf{IssueVC}(\cdot)$, as well as the proving key sizes, increase with the system scale due to the growing circuit complexity, the verification phase demonstrates exceptional efficiency and stability, leveraging the succinctness of the Groth16 protocol. 
Regardless of the increase in VC or Merkle tree size, verification times ($\mathsf{VerifyIss}(\cdot)$ and $\mathsf{VerifyVC}(\cdot)$) remain consistently low at approximately 0.3 seconds, proof sizes ($\boldsymbol{\pi}_{\mathtt{VC}}$ and $\boldsymbol{\pi}_{S}$) stay constant at around 800 bytes, and verification key sizes remain invariant. This strict $O(1)$ verification complexity and communication cost confirm the scheme's high scalability and minimal resource consumption for verification in large-scale scenarios.

 \begin{figure}[!t]
 \centerline{\includegraphics[width=1.0\linewidth]{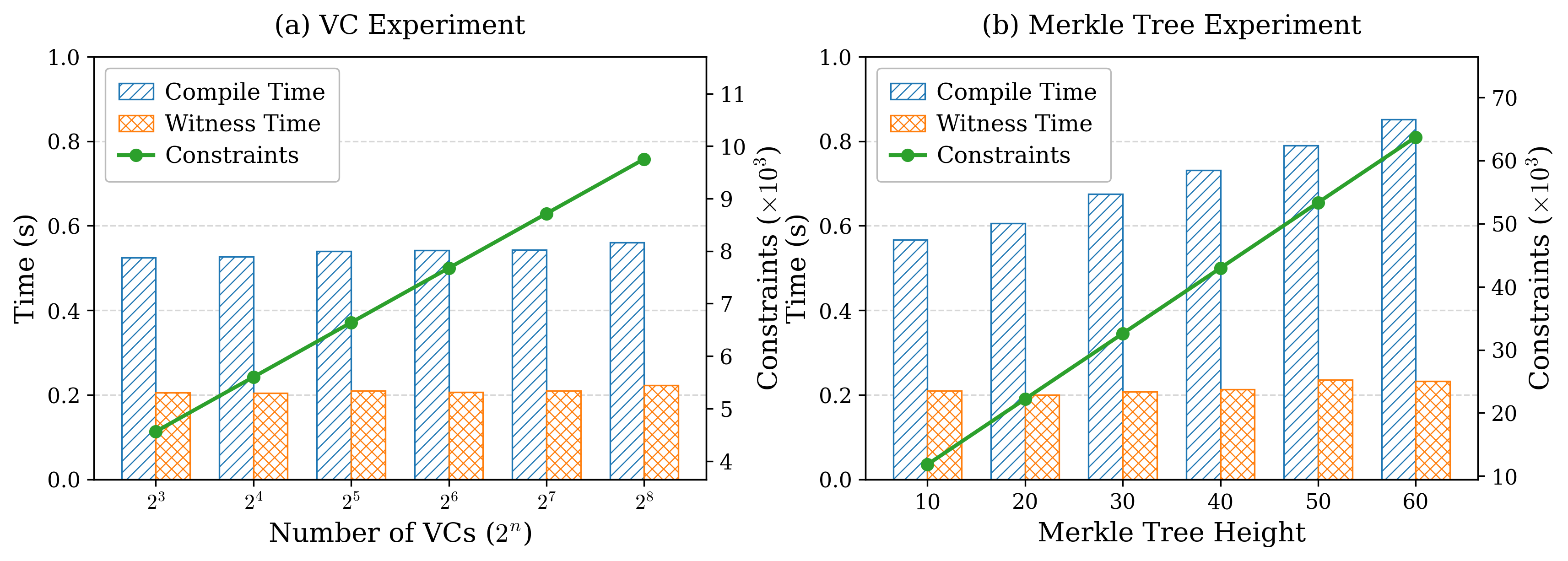}}
 \caption{Performance evaluation of the ZKP circuit for statement $\mathcal{S}$. The results display the compilation time, witness generation time, and the number of circuit constraints under (a) varying numbers of VCs and (b) varying Merkle tree heights.} \label{fig:ex3a}
 \end{figure}

\subsubsection{Experiments for Membership Proofs and Qualification Proofs}

We implement and evaluate the ZKP circuit illustrated in Fig.~\ref{fig:ZK} for the statement $\mathcal{S}$, testing three key metrics to demonstrate the additional off-chain costs introduced by the issuer's qualification proof ($\boldsymbol{\pi}_{S}$). 
Specifically, compile time primarily involves translating the high-level computation logic into an arithmetic circuit and generating the corresponding proving and verification keys, while witness time involves computing the specific assignment of values to all circuit wires that satisfy the circuit's constraints based on the private inputs.
As shown in the performance evaluation in Fig.~\ref{fig:ex3a}, the number of circuit constraints grows linearly with the Merkle tree height, confirming a computational complexity of $O(H_{\mathtt{MK}})$. 
In addition, Figs.~\ref{fig:ex1} and \ref{fig:ex3a} indicate that in addition to the qualification proof ($\boldsymbol{\pi}_{S}$) being extremely succinct, the compilation and witness generation times remain within a practical range even as the number of VCs and the Merkle tree height vary. Specifically, the qualification proof ($\boldsymbol{\pi}_{S}$) consists of only two elements in $\mathbb{G}_1$ and one element in $\mathbb{G}_2$, thereby ensuring that the on-chain verification overhead remains at a minimal constant level.


  \begin{figure*}[!t]
 \centerline{\includegraphics[width=1.0\textwidth]{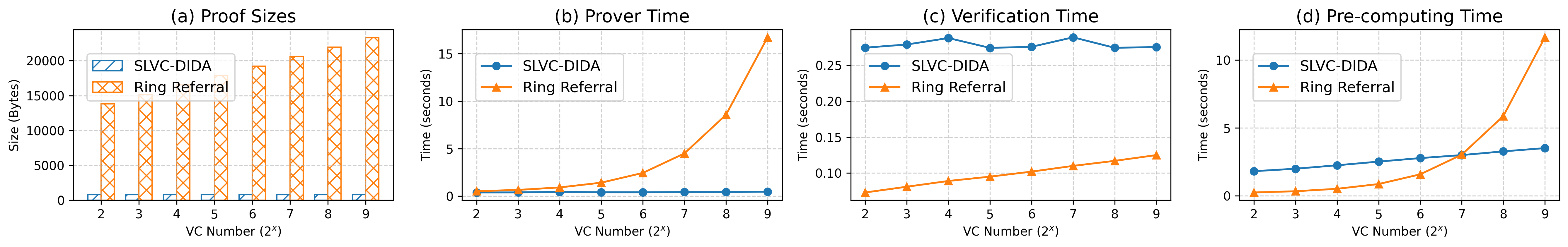}}
 \caption{Performance comparison between SLVC-DIDA and Ring Referral schemes across varying VC numbers. The evaluation metrics include: (a) Proof Sizes, (b) Prover Time, (c) Verification Time, and (d) Pre-computing Time. Notably, the Pre-computing Time of SLVC-DIDA involves the total overhead for circuit compilation, witness generation, and ZKP setup.} \label{fig:ex4}
 \end{figure*}

  \begin{figure*}[!t]
 \centerline{\includegraphics[width=1.0\textwidth]{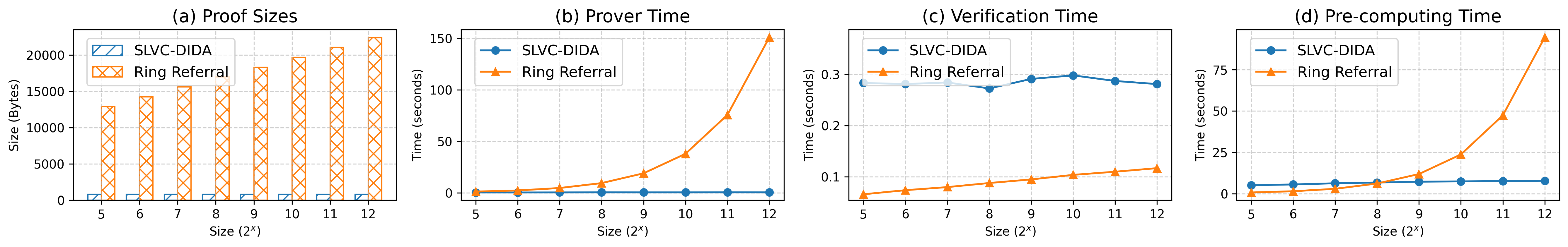}}
 \caption{Performance comparison between SLVC-DIDA and Ring Referral schemes across different sizes. The x-axis represents the Merkle size for SLVC-DIDA or the Ring size for Ring Referral.} \label{fig:ex5}
 \end{figure*}

\textbf{Comparison with ring signature-based VC scheme.}
To further demonstrate the superior performance of SLVC-DIDA in DID authentication, we implement the Ring Referral scheme \cite{ta2025ring}, a credential authentication model based on ring signatures and zk-SNARK, and compared it with SLVC-DIDA across varying VC numbers and Merkle/Ring sizes. 
Fig. \ref{fig:ex4} illustrates the performance comparison across varying VC numbers (from $2^2$ to $2^9$), while Fig. \ref{fig:ex5} depicts the scalability test across different Merkle tree or Ring sizes (increasing from $2^5$ to $2^{12}$). 
The results indicate that SLVC-DIDA performs significantly better in large-scale scenarios. 
Specifically, regarding Prover Time in Fig. \ref{fig:ex5}(b), as the Merkle/Ring size reaches $2^{12}$, Ring Referral's time increases to approximately 150 seconds, whereas SLVC-DIDA remains consistently low at under 1 second. 
Similarly, regarding Proof Sizes in Fig. \ref{fig:ex4}(a), as the VC number increases to $2^9$, Ring Referral's proof size exceeds 20,000 bytes; in contrast, SLVC-DIDA requires only a few hundred bytes and remains constant.

In terms of computational complexity during the proving phase, Ring Referral \cite{ta2025ring} explicitly states that its process involves $3n$ pairings and $2\log n$ group exponentiations, where $n$ represents the ring size. 
This implies a linear prover complexity of $O(n)$, which explains why Ring Referral's proving time exhibits significant growth as $n$ increases in Fig. \ref{fig:ex5}(b). 
In contrast, SLVC-DIDA utilizes a Merkle tree accumulator combined with Groth16-based zk-SNARKs. 
For the prover, the primary overhead consists of generating the Merkle proof path, which has a complexity of $O(\log n)$, and generating a constant-size SNARK proof. 
Since $O(\log n)$ grows slowly relative to $n$, the performance curve for SLVC-DIDA demonstrates a clear efficiency advantage over the linear scaling of Ring Referral.

Regarding pre-computation and setup overhead, Fig. \ref{fig:ex5}(d) shows that Ring Referral's pre-computing time reaches approximately 90 seconds at a size of $2^{12}$. 
This high latency occurs because the model should dynamically compute parameters, such as accumulators or aggregated public keys, based on specific ring members. 
Conversely, SLVC-DIDA's pre-computing time is primarily consumed by circuit compilation and the one-time Trusted Setup required by Groth16. 
Crucially, this overhead is fixed and does not require re-computation as new users join the system. 
The low overheads show that SLVC-DIDA has better scalability in dynamic, high-concurrency DID authentication scenarios.

\begin{table}[!t]
\centering
\caption{On-chain Overhead of SLVC-DIDA}
\label{tab:performance-comparison}
\begin{tabular}{ccccc}
\hline
\textbf{Algorithms}             & \textbf{\begin{tabular}[c]{@{}c@{}}Deploy \\ Gas\end{tabular}} & \textbf{\begin{tabular}[c]{@{}c@{}}Verify \\ Gas\end{tabular}} & \textbf{\begin{tabular}[c]{@{}c@{}}Verify \\ Time (s)\end{tabular}} & \textbf{\begin{tabular}[c]{@{}c@{}}Proof Size \\ (bytes)\end{tabular}} \\ \hline
$\mathsf{VerifyIss}$ ($\boldsymbol{\pi}_{\mathtt{VC}}$) & 393,157                                                        & 221,626                                                        & 0.2784                                                              & 804.5                                                                  \\
$\mathsf{VerifyVC}$ ($\boldsymbol{\pi}_{S}$)            & 419,614                                                        & 229,108                                                        & 0.2680                                                              & 805.4                                                                  \\ \hline
\end{tabular}
\end{table}

\subsubsection{Experiments for Blockchain}

To demonstrate the practicality of SLVC-DIDA in blockchain environments, we evaluated the on-chain overhead of our two core verification algorithms on the Ethereum blockchain. The evaluation metrics include Deploy Gas, which represents the one-time cost required to upload the smart contract to the blockchain, and Verify Gas, which indicates the recurring computational cost for validating proofs on-chain. Notably, the verification costs for both the membership proof ($\boldsymbol{\pi}_{\mathtt{VC}}$) and the qualification proof ($\boldsymbol{\pi}_{S}$) remain constant at $O(1)$, regardless of the number of VCs or the size of the Merkle tree. 
Consequently, the results presented in Table \ref{tab:performance-comparison} are calculated as the average values across a wide range of parameters, specifically varying the number of VCs ($N_{\mathtt{VC}}$) from $2^2$ to $2^9$ and the Merkle tree height ($H_{\mathtt{MK}}$) across depths of $\{ 10, 20, 30, 40,50 \}$.

The detailed experimental results in Table \ref{tab:performance-comparison} illustrates the efficiency of the SLVC-DIDA on the Ethereum. 
For the $\boldsymbol{\pi}_{\mathtt{VC}}$ in algorithm $\mathsf{VerifyIss}(\cdot)$, the one-time deployment cost is 393,157 Gas, while the recurring verification cost is 221,626 Gas, with an average verification time of approximately 0.2784 seconds and a proof size of 804.5 bytes. 
Similarly, the VC verification algorithm $\mathsf{VerifyVC}(\cdot)$ for $\boldsymbol{\pi}{S}$ incurs a deployment cost of 419,614 Gas and a verification cost of 229,108 Gas. 
Therefore, Table \ref{tab:performance-comparison} confirms that both the storage costs (proof size) and computational costs (Gas and time) are kept limited, validating that SLVC-DIDA is highly efficient and suitable for deployment on resource-constrained blockchain networks.

\section{Conclusion}
\label{sec:coc}
In this paper we have proposed SLVC-DIDA, the first issuer-hiding and privacy-preserving DID authentication scheme that implements signature-less VC.
SLVC-DIDA uses the hash value of nonce to establish a link between VCs and the issuers, and substitutes the signature key in PKI with a ZKP of the issuer's legitimate qualifications. 
Furthermore, SLVC-DIDA ensures issuer-hiding by maintaining zero-knowledge and correctness during VC issuance and verification through ZKP with Merkle accumulator.
A Merkle SNARK-based VC list is stored on the blockchain, enabling the public verifiability of DIDs and VCs. 


\bibliographystyle{IEEEtran}
\bibliography{jobname}

\vfill

\end{document}